# A comparison of the impacts of CMEs and CIRs on the Martian dayside and nightside ionospheric species

L. Ram[1], D. Rout[2], R. Rathi[1], S. Mondal[1,†], S. Sarkhel[1*], and J. Halekas[3]

[*]Sumanta Sarkhel, Department of Physics, Indian Institute of Technology Roorkee, Roorkee - 247667, Uttarakhand, India (sarkhel@ph.iitr.ac.in)

[1]Department of Physics,
Indian Institute of Technology Roorkee,
Roorkee - 247667
Uttarakhand, India

[2]GFZ German Research Centre for Geosciences,
Potsdam, Germany

[3]Department of Physics and Astronomy
414 Van Allen Hall
University of Iowa
Iowa City, IA – 52242
USA

[†]Now at: Physical Research Laboratory,
Ahmedabad-380009,
Gujarat, India

**Abstract**

Measurements from the Mars Atmosphere and Volatile EvolutioN (MAVEN) spacecraft, orbiting Mars are used for investigating the impact of coronal mass ejections (CMEs) and corotating interaction regions (CIRs) on Martian ionospheric species. We have chosen 15 CME and 15 CIR events (2015-2020) at Mars from the existing catalogs. We have extensively analyzed the Martian dayside and nightside profiles of ionospheric species during each of the CME and CIR events. We have selected those orbit plasma density profiles which showed significant differences from the mean quiet-time profile during each event. The primary focus of this paper is to provide a comparative average scenario of the variation of Martian ionospheric species during CMEs and CIRs events. A significant difference can be observed in the profiles of the Martian dayside and nightside ionospheric species ($O^+$, $O_2^+$, $CO_2^+$, $NO^+$, $C^+$, $N^+$, & $OH^+$) during CMEs and CIRs in comparison to mean quiet-time profile. The difference is more prominent on the nightside compared to the dayside ionosphere. During CIRs, the nightside ion density is nearly one order of magnitude less (above 250 km) in comparison to CMEs. The mean peak altitude and density of the lighter ions ($O^+$, $C^+$, $N^+$, & $OH^+$) were at lower altitudes during the CIRs compared to CMEs. Therefore, this study suggests that during the declining phase of solar cycle (SC 24), the impact of CIRs on the Martian ionospheric species is more prominent compared to CMEs.

**Key Points**

1. Martian ionospheric species show noticeable depletion during CMEs/CIRs compared to mean quiet-time density.
2. During CIRs, nightside mean ion density above 250 km is one order less in magnitude compared to CMEs.
3. The mean peak altitude and density for lighter ions are lower on both dayside and nightside during CIRs compared to CMEs.

## Plain language summary

The space weather events like coronal mass ejections and corotating interaction regions pose significant effects on the ionosphere of Mars. This motivated us to study the behavior of the Martian dayside and nightside ionosphere and the potential role of these events. The primary focus of this study is to provide a comparative average scenario for ionospheric species variation during these two events. The ionospheric average ion density profiles have notable differences during both events. During corotating interaction regions, the average plasma density depletion is more prominent on the nightside as compared to the dayside. The nightside average ion density profiles have one to two orders less magnitude (>250 km) with lower mean peak altitude (for lighter ions $O^+$, $C^+$, $N^+$, and $OH^+$) during corotating interaction regions in comparison to both coronal mass ejections and mean quiet-time profile. The average analyses indicate that corotating interaction regions have a greater impact on the Martian ionosphere in comparison to coronal mass ejections during the declining phase of the solar cycle 24. Therefore, these results will enhance our understanding of the Martian dayside and nightside upper ionosphere and the potential role of these events in the evolution of planetary atmospheres over time.

## 1. Introduction

Solar energetic events, namely Coronal Mass Ejections (CMEs) and Stream Interaction regions /Corotating Interaction Regions (SIRs/CIRs), are the principal drivers of the major space weather phenomena in the planetary atmosphere. CMEs are defined as a huge burst of plasma and magnetic field originating from the outer surface of the sun (Gosling, 1990; Gopalswamy, 2006; Jian et al., 2006). CMEs traverse all the way from the solar corona to interplanetary space, preceded by a forward shock (Chen, 2011; Tousey, 1973). They are commonly characterized by a strong rotating magnetic field, low β (ratio of plasma pressure and magnetic pressure), low ion temperature, high proton density, and enhancement in velocity (Cane & Richardson, 2003; Gosling et al., 1991; Gosling & Forsyth, 2001; Neugebauer & Goldstein, 1997; Russell & Shinde, 2005; Wimmer-Schweingruber et al., 2006; Zurbuchen & Richardson, 2006).

During the declining phase of the solar cycle, the high-speed solar wind originates from the coronal holes (at higher solar latitudes) and the slow solar wind streams originate at lower solar latitudes (Belcher & Davis Jr., 1971; Borovsky & Denton, 2006; Gosling & Pizzo, 1999; Richardson, 2018; Smith & Wolfe, 1976). The high-speed solar wind interacts with the preceding slow solar wind streams to form an interaction region called a Stream Interaction regions (SIRs) (Pizzo, 1978; Richardson, 2018). Further, if the source of the SIRs co-rotates once around the solar axis, it is referred as Corotating Interaction Regions (CIRs) (Dubinin et al., 2009; Rout et al., 2017; Wei et al., 2012). The boundary that separates the high-speed streams from the slow streams is called the stream interface (SI). It is associated with an increase in the proton temperature and a high-pressure region (bounded by forward and reverse shocks) at large heliospheric distances (Burlaga, 1974). While this paper includes both SIRs and CIRs, we employ the terminology CIRs throughout for simplicity.

CMEs and CIRs drive the major space weather events on the Earth and other planets (like Venus and Mars) of our solar system. On Earth, they initiate various geomagnetic events viz. storms and substorms. The Earth enjoys the advantage of a global magnetic shield, whereas Mars lacks such a global intrinsic magnetic shield, which consequently allows the solar wind to interact directly with the Martian atmosphere (Cloutier et al., 1969; Connerney et al., 2015). A Martian bow shock is formed as an obstacle in the form of a shock boundary where the normal component of the supersonic solar wind decreases to subsonic speeds. Inside the shock boundary, there is a region of shocked and turbulent solar wind plasma termed the magnetosheath (Dubinin et al., 1997; Lundin et al., 1990). At the lower boundary of that region, the shocked solar wind plasma interacts with the upper atmosphere of Mars, particularly in the ionosphere.

The dayside ionosphere is formed via photo-ionization by the solar extreme ultraviolet (EUV) radiation (Fox & Dalgarno, 1979; McElroy et al., 1977). The current induced in the ionosphere by the motional electric field of the solar wind plays a crucial role in the formation of the induced magnetosphere (Michel, 1971). A tangential discontinuity in thermal plasma density that marks the transition from the hot plasma in the induced magnetosphere to the cold dense ionospheric plasma is termed as ionopause (Chu et al., 2021; Duru et al., 2020; Vogt et al., 2015). Thus, an ionopause is identified as the sudden decrease in the electron density and an enhancement in the electron temperature (Chu et al., 2021; Jethwa, 2021; Krishnaprasad et al., 2019; Sánchez-Cano et al., 2020; Thampi et al., 2021; Vogt et al., 2015). The Martian ionopause altitude is affected by various space weather events (Krishnaprasad et al., 2019, 2021; Schunk & Nagy, 2009; Thampi et al., 2018, 2021). Using MAVEN observations, Thampi et al. (2018) showed the depletion of $O^+$, $O_2^+$, and $CO_2^+$ ions during the CME event and reported the lowering of ionopause during the passage of two consecutive CMEs.

There are various studies that reported the impact of space weather events on Mars. The Martian upper atmosphere responds to the enhancement in the solar wind density, velocity, dynamic pressure, and the variable interplanetary magnetic field (IMF). For example, the solar wind interaction with the Martian upper atmosphere leads to the escape of various ions and neutral species. The escape of oxygen ions and precipitation of heavy ions in the Martian magnetosphere have been reported using data from ASPERA-3/Mars Express (MEX) (Hara et al., 2011; Wei et al., 2012). Further, the work of Krishnaprasad et al. (2019) found enhanced pickup ion-induced depletion of the nightside of the Martian ionosphere and depletion of electron density observed for a longer period of time. The electron temperature was also enhanced during the period of electron density depletion. In addition, Opgenoorth et al. (2013) studied the signature of increased plasma transport over the terminator and enhanced ion escape from the upper ionosphere of Mars during CMEs and CIRs events. The study by Jakosky et al. (2015) reported the enhancement in the escape rate of Martian ions to space during the major impact of CME using MAVEN observations. Lee et al. (2017) reported heating of the upper atmosphere by solar flare emissions and enhanced atmospheric escape rates. Ma & Nagy, (2007) studied the ion escape fluxes from Mars using a 3D, multi-species ($O^+$, $O_2^+$, $CO_2^+$, and $H^+$), non-ideal MHD model. Several studies during variable solar wind conditions, such as CIRs and CMEs have also demonstrated the depletion and energization of the Martian ionosphere during these solar events (Benna et al., 2015; Huang et al., 2019; Ma & Nagy, 2007; Richardson, 2018; Thampi et al., 2021; Withers et al., 2018). However, there is no comprehensive investigation that shows the relative difference between the impact of CMEs and CIRs on the Martian ionosphere to the best of our knowledge. Further, this study, for the first time, also reports the effects on the ionospheric species (both the heavier ions: $O_2^+$, $CO_2^+$ & $NO^+$ and the lighter ions: $O^+$, $C^+$, $N^+$, & $OH^+$), on the dayside and nightside during CME and CIR events. In the present investigation, we performed a detailed comparative analysis by considering 15 CMEs and 15 CIRs using multiple MAVEN datasets. We find more depletion in the plasma density for Martian dayside and nightside ionosphere during CIRs in comparison to CMEs. This paper is organized as: Section 2 gives a description of the datasets used for this study, Section 3 presents the results, Section 4 provides a discussion on the depletion of ionospheric plasma, and Section 5 concludes the paper.

## 2. Data

In order to study the Martian ionosphere and solar wind environment, we utilized multiple instruments onboard the MAVEN spacecraft. It was launched in November 2013, with orbital insertion occurring in September 2014, with its primary science mission beginning in November 2014. With an apoapsis of 6200 km and periapsis of 150 km, it is capable of measuring the upstream solar wind before it is affected by the Martian environment. The solar wind parameters (density, velocity, and dynamic pressure) and the IMF near Mars are obtained from the Solar Wind Ion Analyzer (SWIA) and Magnetometer (MAG) instruments aboard the MAVEN spacecraft. SWIA measures the solar wind ion flow around Mars in the upstream solar wind, magneto-sheath, and tail regions inside the bow shock. SWIA covers a broad energy range of 5 eV to 25 keV for solar ions (Halekas et al., 2015). The IMF is measured from the Magnetometer (MAG) instrument. It has two independent triaxial fluxgate magnetometer sensors for magnetic field investigation and has an intrinsic sample rate of 32 vector samples/second (Connerney et al., 2015). The Level 2, version_01, revision number _01 (v01_r01) data of SWIA (Halekas, 2017) and the Level 2, version_01, revision number _01 (v01_r01) data of MAG (Connerney, 2017) are used for upstream solar wind observations. The electron density and electron temperature data are obtained using the Langmuir Probe and Wave (LPW) instrument aboard the MAVEN spacecraft. LPW is optimized to measure electron density in the range of ~$10^2$ $cm^{-3}$ to $10^6$ $cm^{-3}$ and temperature in a broad range of 500-50000 K (Andersson et al., 2015). The ions ($O^+$, $O_2^+$, $CO_2^+$, $NO^+$, $C^+$, $N^+$, & $OH^+$) density data is obtained using the Neutral Gas and Ion Mass Spectrometer (NGIMS) aboard the MAVEN spacecraft. NGIMS is designed to measure the ions and neutrals in the range of 125-500 km altitude region of the Martian atmosphere using a dual ion source and a quadrupole mass analyzer (Mahaffy et al., 2015). It covers the mass range between 2-150 amu with unit mass resolution. The Level 2, version_02, revision_01 (v02_r01) data of LPW (Andersson, 2017) and the Level-2, version_v08, revision number_r01 (v08_r01) data of NGIMS (Benna & Lyness, 2014) are used in our analyses. Both LPW and NGIMS measure on the dayside and nightside during the periapsis (150-500 km above the Mars surface) (Andersson et al., 2015; Mahaffy et al., 2015). All the datasets were accessed through Python Data Analysis and Visualization tool (PyDIVIDE) as well as the Planetary Data System (PDS).

In order to select the 15 CME events, we have utilized the catalogs of already identified CME events provided by Zhao et al. (2021) and the Space Weather Database of Notifications, Knowledge, and Information (DONKI) system. Similarly, the 15 CIR events have been selected from the catalogs provided by Geyer et al. (2021) and Huang et al. (2019) for our analyses. We have also utilized the Wang-Sheelay-Arge (WSA)-ENLIL+Cone model from the ENLIL Solar Wind Prediction to track the planetary positions, plasma feature, compressed streams, and IMF over the heliosphere (Mays et al., 2015; Odstrcil, 2003).

## 3. Results

In order to study the impacts of CMEs and CIRs on the Martian dayside and nightside ionospheric species, we have analyzed the ion density profiles during these events. The following subsections describe the behavior of the Martian ionosphere during specific cases of CME and CIR events. In the last subsection, we have provided a comparative average scenario of the Martian dayside and nightside ionospheric species variation during 15 CMEs and 15 CIRs.

### 3.1 Identification of passage of the CME and CIR events at Mars:

The period from 03-05 March 2015 on Mars was characterized by a CME event and the period from 19-21 March 2018 was characterized by a CIR event. Figure 1 shows the WSA-ENLIL+Cone simulation snapshots in the ecliptic plane during specific CME (03-05 March 2015) and CIR (19-21 March 2018) events. The color contour represents the radial solar wind velocity whereas the black-white dashed line represents the IMF in the inner heliosphere. The position of the planet Mars along with other inner planets is shown in the figure. Figures 1a-c show the three stages of development for the CME structure while entering, interacting, and leaving Mars. The CME event expands and becomes wider as it propagates through the heliosphere. During the event, a portion of the CME blob (pointed by the arrow in red color in Figures 1a-c) hit the planet Mars. Similarly, Figures 1d-f represent the three development stages during the CIR. The green and cyan color contours show the velocity of the solar wind and indicate the stream interaction region (pointed by the arrow in red color in Figures 1d-f) of the slow and fast solar wind near Mars. The ENLIL+Cone simulation predicts the event arrival and departure at Mars during the period.

During these periods for both the (CME [03-05 March 2015] and CIR [19-21 March 2018]) events, MAVEN observed that the solar wind parameters changed abruptly while crossing the Martian bow shock. In order to avoid contamination from the measurements near the bow shock and to select the undisturbed solar wind intervals in the upstream region, we utilized an algorithm proposed by Halekas et al. (2017). We tune this algorithm with $|v| > 200$ km/s (bulk flow speed), $\sigma_B/|B| < 0.15$ (normalized magnetic field fluctuation levels) and $R > 4200$ km (altitude) reliably lie in the undisturbed solar wind. Figure 2 shows the variation of upstream solar wind (a) density, (b) velocity, (c) dynamic pressure, and (d) IMF (|B|) during 01-06 March 2015 CME event. The enhancement in the solar wind density started on 03 March 2015 and attained the peak of ~30 $cm^{-3}$. The solar wind velocity showed an enhancement starting from 03 March 2015, and the highest velocity ~500 km/s was observed on 04 March 2015 and lasted until early 05 March 2015. This event was also associated with enhancements in the dynamic pressure and IMF (|B|). The peak dynamic pressure and the peak IMF (|B|) were ~9 nPa and ~20 nT, respectively on 03-04 March 2015. The vertical dotted lines (in different colors) from left to right in each panel indicate the MAVEN orbits 822 (blue), 823 (cyan), 824 (magenta), 825 (red), and 826 (brown) during the disturb time. The short horizontal lines in the time series plot indicate the unobserved period as SWIA data set covers ~60% of the MAVEN orbits primarily on the upstream region (Halekas et al., 2017).

Similarly, Figure 3 shows the variation of upstream solar wind (a) density, (b) velocity, (c) dynamic pressure, and (d) IMF (|B|) during 17-24 March 2018 CIR event. The solar wind velocity showed an enhancement starting from 20 March 2018 and attained the peak of ~594 km/s and also sustained a higher magnitude for several days. The enhancement in the solar wind density started on 20 March 2018 and attained the peak of ~19 $cm^{-3}$. The peak dynamic pressure and the peak resultant IMF (|B|) were ~6 nPa and ~15.8 nT, respectively on 20 March 2018. The vertical-colored dotted lines from left to right in each panel indicate the MAVEN orbits 6748 (blue), 6749 (cyan), 6750 (magenta), 6751 (red), and 6752 (brown) during the disturb time. Since the MAVEN spacecraft follows an elliptical orbit with an inclination angle of ~75°, it observes the upstream solar wind measurements only intermittently, making it difficult to observe the exact event arrival time at Mars (Halekas et al., 2017).

### 3.2 Impact of the CME on the Martian ionosphere during 03-05 March 2015 event:

In order to study the impact of the incoming CME on the Martian ionosphere, we have analyzed the LPW and NGIMS data during CME (03-05 March 2015) event. The LPW and NGIMS periapsis orbit measurements during the enhancement in solar wind conditions are marked with vertical-colored dotted lines (shown in Figure 2). The colors used for depleted density profiles (Figures 4 & 5) are identical as used for vertical-colored dotted lines in Figure 2. During 03-05 March 2015 event, the MAVEN inbound phase monitors the dayside (SZA ~53°-84°), and the outbound phase monitors the nightside (SZA ~84°-115°) of Mars. Figures 4a-d show the LPW periapsis observations of electron density and temperature for dayside and nightside of the Martian ionosphere by MAVEN orbits. Figures 4a-b (top panel) and Figures 4c-d (bottom panel) show electron density and temperature profiles on the dayside and nightside, respectively. The mean quiet-time profile is shown with a black solid line and calculated by obtaining the mean of quiet orbits before the event (02-03 March 2015, 5 quiet orbit profiles with adequate data) with a standard deviation (STDEV). The enhancement in the upstream solar wind parameters was observed on 03 March 2015 around 17:22 UTC during the passage of CME and the disturb period lasted until early 05 March 2015 around 03:02 UTC (Figure 2). The MAVEN orbits during the disturbed period were 822 (17:22 UTC), 823 (18:52 UTC), 824 (02:20 UTC), 825 (06:50 UTC) and 826 (11:21 UTC). The enhancement in the solar wind conditions leads to the depletion in the electron density compared to the mean quiet-time profile. The event time orbit profiles are depicted in different colors as shown in the Figure 4. The dayside profiles show that during 03-04 March 2015, the electron density was notably reduced in orbits 822 (above 250 km), 823 (above 300 km) and 826 (above 270 km) associated with an enhancement in temperature up to ~10000 K. The sharp decrease in the ion density profile signifies the ionopause-like feature. Similarly, the nightside profiles (orbits 823 and 824 on 03-04 March 2015) showed deviation beyond the STDEV of mean quiet-time profile. In orbits 823 and 824, a clear deviation from the mean quiet-time profile and sharp density gradient can be observed nearly at 220 km for both the orbit profiles (similar to the ionopause-like features). The electron density was depleted completely at around 400 km during the 823 orbit profile and at around 300 km during the 824 orbit profile. An enhancement of electron temperature was also observed which approached more than 10000 K for the orbits 823 and

824. These electron density profiles show a greater variation on the nightside than on the dayside after a CME impact. The ionopause-like features form at lower altitude in the nightside (~220 km) in comparison to the dayside (~ 250-300 km).

In addition to electron density and temperature, it is also important to explore the effects of the CME on various ions. Figure 5 shows the NGIMS observation for various ions ($O^+$, $O_2^+$, $CO_2^+$, $NO^+$, $C^+$, $N^+$, & $OH^+$) density for dayside (top panel of Figures 5a-g) and nightside (bottom panel of Figures 5h-n) of the Martian ionosphere measured by the MAVEN, respectively. The enhancement in the solar wind conditions leads to the depletion in the ion density compared to the mean quiet-time profile. During orbits 822, 823, 825 and 826 (shown by vertical dotted lines in different color in Figure 2), the deviation in ion density profiles is observed on both dayside and nightside. On the dayside, the $O^+$ density profile (Figure 5a) shows a clear deviation from the mean quiet-time profile for the orbits 822 and 826. The $O^+$ density shows a sharp density gradient starting from 250 km for the orbit 822 and nearly 270 km for the orbit 826, respectively. This situation is similar to the ionopause-like feature as defined by Vogt et al. (2015). Figure 5b shows a similar trend for $O_2^+$ ion density during the same orbits as observed for $O^+$. The ion density profiles for other ions ($CO_2^+$, $NO^+$, $C^+$, $N^+$, & $OH^+$) also depict similar variations for the corresponding orbits as shown for $O^+$ and $O_2^+$. Therefore, during the dayside, the ionopause-like features are clearly observed from the electron and ion density profiles for orbits 822 and 826 in both LPW and NGIMS measurements. This indicates that there were evidently two orbit profiles in the dayside which depict a sharp density gradient (around 250 km for orbit 822 and 270 km for orbit 826) and had an ionopause formation around 450 km altitude.

During the nightside (Figures 5h-n), there was a clear deviation from the mean quiet-time profile for each ion and a sharp density gradient for $O^+$, $O_2^+$, $CO_2^+$, and $NO^+$ ions can be observed for the orbits 823 and 825. Here, the sharp density gradient observed for the two orbit profiles was at a relatively lower altitude (~220 km and ~260 km) as compared to the dayside (around 250 km for the orbit 822 and 270 km for the orbit 826). For the remaining ions ($C^+$, $N^+$, & $OH^+$), the corresponding orbit profile also showed the deviation from mean quiet-time. The depleted density profile is steeper for orbit 823 for all ions in comparison to orbit 825. The ionopause altitude form nearly 400 km during this orbit 823. Therefore, during the nightside, the ionopause-like features are clearly observed from the electron and ion density profiles for orbit 823 in both LPW and NGIMS measurements. The ionopause altitude observed at lower altitude in the nightside (~400 km) as compared to the dayside (~450 km). Furthermore, these observations for ions ($O^+$, $O_2^+$, $CO_2^+$, $NO^+$, $C^+$, $N^+$, & $OH^+$) indicate high depletion with lower ions densities and formation of ionopause-like feature during orbit 823 in the nightside of the Martian ionosphere during this CME event.

### 3.3 Impact of the CIR on the Martian ionosphere during 19-21 March 2018 event:

In order to study the impact of the incoming CIR on the Martian ionosphere, we have analyzed the LPW and NGIMS data during a CIR (19-21 March 2018) event. The LPW and NGIMS periapsis measurements during the enhancement in solar wind conditions are marked with vertical dotted lines in different colors (shown in Figure 3). The colors used for depleted density profiles (Figures 6 & 7) are identical as used for vertical dotted lines in Figure 3. During this event, the MAVEN inbound and outbound phases monitor the dayside (SZA ~61°-96°) and nightside (SZA ~112°-144°) of Mars, respectively. Figures 6a-d show the LPW periapsis observation of electron density and temperature during dayside and nightside of MAVEN orbits. Figures 6a-b (top panel) and Figures 6c-d (bottom panel) show the electron density and temperature profiles on the dayside and nightside, respectively. The mean quiet-time profile is shown with a black solid line and calculated by taking the mean of quiet orbit profiles (5 quiet orbit profiles with adequate data) with STDEV during17-18 March 2018. The event time orbits (which started on 19 March and lasted up to 21 March 2018) are shown with different colors. Here, we have shown the disturbed time orbits. The enhancement in the upstream solar wind parameters was observed on 19 March 2018 around 23:18 UTC during the passage of CIR and disturb period remain till 21 March 2018, 04:18 UTC (Figure 3). The MAVEN orbits during the disturbed period were 6748-6752. The plasma depletion observed in the ionosphere during these orbit profiles. The dayside density profiles (Figures 6a-b) show that during the CIR event, the electron density for orbits 6748-6751 (20 March 2018) was notably reduced with an enhancement in the temperature up to ~10000 K compared to mean quiet-time profile. In addition, we have observed a sharp density gradient (~250 km) during the orbits 6748 and 6749, where the density is nearly depleted above 350 km. Similarly, the nightside profiles (Figures 6c-d) with orbits 6750-6752 (20 March 2018) show deviation in the electron density beyond the STDEV of the mean quiet-time profile. During the orbit 6750 (12:44 UTC), the electron density was depleted above 300 km altitude. The orbits 6751 (17:14 UTC) and 6752 (21:41 UTC) are most depleted, however there is very less data available for these orbits to confirm the depletion. An enhancement of electron temperature can also be observed, reaching over 10000 K for the 6750-6752 orbits. The depletion in the electron density with the enhancement in the temperature indicates the formation of an ionopause.

Figure 7 shows ion density profiles measured by NGIMS on the Martian dayside (Figures 7a-g) and nightside (Figures 7h-n). The dayside $O^+$ density profile (Figure 7a) corresponding to the orbits 6748 (03:48 UTC) and 6749 (08:15 UTC) during disturbed solar wind conditions show the deviation from the mean quiet-time profile with depletion from 100 $cm^{-3}$ to 0.01$cm^{-3}$ above 350 km altitude. A similar trend is observed for $O_2^+$, $CO_2^+$, $NO^+$, $C^+$, $N^+$, & $OH^+$ (Figures 7b-g). A sharp density gradient starting (~250 km) is observed during the orbits 6748 and 6749, indicating the formation of ionopause-like features and ionopause formed around 350 km altitude. Therefore, during the dayside, the ionopause-like features are clearly observed in the electron and ion density profiles for orbits 6748 and 6749 in both LPW and NGIMS measurements. This indicates that there were evidently two orbit profiles in the dayside which depict a sharp density gradient (around 250 km for orbits 6748 and 6749) and had an ionopause formation at around 350 km altitude.

On the nightside (Figures 7h-n), a clear deviation from the mean quiet density profile is observed for all ions. A sharp density gradient is observed near 200 km for orbits 6750 and 6751, indicating the formation of an ionopause. The altitude of the nightside ionopause is much lower than the dayside, which formed near 300 km, highlighting the highly variable nature of the nightside ionosphere. Although there is very less data available for ion species, so we cannot infer the exact altitude for sharp density gradient in the nightside. However, both LPW and NGIMS measurements show the depletion during orbits 6750 and 6751. These measurements depicted that the electron and ion density depletions occurred at lower altitude on the nightside compared to the dayside. This indicates that the Martian ionosphere was highly variable and depleted during the nightside compared to the dayside.

### 3.4 Comparison of the impact of CMEs and CIRs on the Martian ionosphere

In the previous sections, we presented specific cases of the ionospheric electron and ion density profiles, which showed the deviation from mean quiet-time during the passage of CME and CIR events in the Martian dayside and nightside ionosphere. In this section, we have performed a comparative average analysis of the effect of several CME and CIR events on the Martian ionosphere. For this, we have extensively analyzed 15 events each. The list of CMEs and CIRs is provided in Table 1, which contains the information about local solar time (LST) and solar zenith angle (SZA) in the Martian dayside and nightside periapsis passes of MAVEN spacecraft during 15 CMEs and 15 CIRs. The mean solar wind density, velocity, dynamic pressure, and IMF for each CME and CIR events are provided in Table 3 as supplementary material. By critically analyzing those 15 CMEs and 15 CIRs, we have selected highly disturbed orbit profiles (which are well beyond the STDEV) for 15 CME events (53 dayside and 55 nightside disturbed orbit profiles) and 15 CIR events (55 dayside and 54 nightside disturbed orbit profiles). For mean quiet-time ionospheric variations, we have selected one quiet-time orbit profile before each CME and CIR event. As a result, the mean dayside (nightside) quiet-time profile is generated by taking the average of 30 quiet-time dayside (nightside) orbit profiles.

In order to address the differences (if any) between CME and CIR events in terms of their ionospheric impact and species variation in the Martian ionosphere, we have explored the variations of electron density, temperature, and ion density. We have calculated the mean of the plasma density with STDEV for all major disturbed orbit profiles separately for all CMEs and CIRs. Figures 8a-d show the collective average profiles for electron density and temperature using LPW periapsis measurements. Figures 8a-b (top panel) and 8c-d (bottom panel) represent electron density and temperature dayside and nightside profiles, respectively. The black color profile represents the mean quiet-time electron density and temperature for both dayside and nightside, while the black shaded area indicates the STDEV. The red color profile represents the mean electron density and temperature for 53 dayside and 55 nightside disturbed orbits during 15 CME events, while the red shaded area indicates the STDEV. The same analysis has been carried out for 55 dayside and 54 nightside disturbed orbits during 15 CIR events, where the mean profile is represented by the blue color profiles and the blue shaded area representing the STDEV. Figures 8a-b show a clear deviation of the dayside electron density and temperature profiles during CMEs and CIRs from the mean quiet-time electron density and temperature profiles. Further, a significant difference between the dayside mean electron density during CIRs versus CMEs with the difference between the two profiles increasing with altitude. On the nightside, the electron density profiles (during CMEs and CIRs) show more deviation from the mean quiet-time profile and deviation increases with altitude. The difference between the mean electron density profiles during CIRs versus CMEs is even more pronounced (Figure 8c) compared to the dayside. The mean electron density during CIRs reached less than 100 $cm^{-3}$ above 350 km, whereas it remained more than 100 $cm^{-3}$ during CMEs. While the mean nightside electron density profile shows a larger variation than the dayside, the difference in the mean temperature profiles is not significant and lies within the STDEV of each other. In general, the electron density at all altitudes is found to be more depleted during CIRs as compared to CMEs. Further, the depletion is more pronounced in the case of nightside than dayside.

Using the same orbits considered in the LPW study, we use the NGIMS observations to investigate the density variation of seven major ions during CME and CIR passages. We selected four lighter ions ($O^+$, $C^+$, $N^+$, & $OH^+$) and three heavier ions ($O_2^+$, $CO_2^+$, & $NO^+$). Figures 9a-g show a clear deviation of the dayside ion density profiles during CMEs and CIRs from the mean quiet-time ion density profiles. A noticeable difference is observed between dayside $O^+$ density profiles during CMEs (red) and CIRs (blue) disturbed orbits (Figure 9a). Ion density began to diverge above 240 km, with CIR profiles showing more than an order of magnitude decrease in the $O^+$ above 450 km compared to CMEs. Similar behavior is observed for the lighter ions $C^+$, $N^+$, and $OH^+$ on both the dayside (Figures 9e-g) and nightside (Figures 9l-n), with the nightside exhibiting a larger decrease in the density. However, the heavier ions ($O_2^+$, $CO_2^+$, & $NO^+$) show a much more pronounced decrease in density at all altitudes during the presence of CIRs as compared to CMEs, particularly on the nightside (Figures 9b-d and Figures 9i-k). The difference between the nightside plasma density during CIRs and mean quiet-time profiles is ranging from one to two orders, whereas it is nearly one order during CMEs.

The mean peak altitude of lighter ions is observed to be lesser during CIR and CME events as compared to mean quiet-time on both dayside and nightside. In addition, the mean peak altitude of lighter ions is observed to be lesser during CIR events as compared to CME events. The peak altitude and associated ion density with STDEV for both dayside and nightside profiles (CMEs vs CIRs) are listed in Table 2. On the dayside, the difference in the mean peak altitude (during CIRs) for $O^+$, $C^+$, $N^+$, and $OH^+$ ranges from 5 to 30 km below that of observed during CMEs. On the nightside, this difference increases, ranging from 10 to 45 km below the peak altitude observed during CMEs.

## 4. Discussions

In the present investigation, we presented a comparative study of the differential impact of the CMEs and CIRs on the Martian ionosphere. The CME (03-05 March 2015) with a solar wind velocity of ~500 km/s and peak dynamic pressure of ~9 nPa had efficiently impacted the topside ionosphere of Mars. This event was a slow CME and the impact is more pronounced on the nightside in comparison to the dayside (Figures 4 & 5). The electron density decreased lower than 100 $cm^{-3}$ together with an enhancement in the electron temperature (~10000 K), forming an ionopause-like feature. A similar observation was reported by Cravens et al. (1982) on the Venusian nightside ionosphere that the depleted electron densities are accompanied by an enhancement in electron temperatures. However, the electron density below 100 $cm^{-3}$ can increase the errors in the LP temperature measurements (Andersson et al., 2015). The observed ionopause-like features for orbits 822, 823, and 826 (~250-270 km) on the dayside are primarily due to the increase of dynamic pressure during this time. On the nightside, the ionopause-like feature observed for the orbit 823 (~220 km) is comparatively at a lower altitude as compared to the dayside. Similarly, during the CIR event (19-21 March 2018), we have observed ionopause-like features on both dayside and nightside. The observed ionopause-like features for orbits 6748 and 6749 started at a height (~250 km) and on the nightside, for the orbits 6750 and 6751, it occurred at ~200 km primarily due to higher dynamic pressure during this time (Figures 6 & 7). These similar ionopause-like features were also reported by previous studies during CME and CIR events (Chu et al., 2021; Jethwa, 2021; Krishnaprasad et al., 2019; Sánchez-Cano et al., 2020; Thampi et al., 2021; Vogt et al., 2015).

During specific cases of CME and CIR, we analyzed seven ion ($O^+$, $O_2^+$, $CO_2^+$, $NO^+$, $C^+$, $N^+$, & $OH^+$) density profiles. The depletion in the ion density has been observed for the same corresponding orbits as observed for electron density during both specific cases of CME (Figure 5) and CIR (Figure 7). The ionopause forms at a higher altitude (~450 km) on the dayside compared to the nightside (~400 km) during CME. Thampi et al. (2018, 2021) reported the depletion in the topside plasma density ($O^+$ and $O_2^+$) with electron temperature enhancement during the CME event. In our study, the depletion is also observed for the other ions. Further, for the CIR case, ionopause forms around ~350 km on the dayside and ~300 km on the nightside. The depleted ion profiles were observed during the disturbed solar wind conditions. According to the previous studies, the enhancement in solar wind dynamic pressure is responsible for a higher depletion of the global ionospheric plasma (Girazian et al., 2017, 2019; Niu et al., 2021). Krishnaprasad et al. (2019) also reported the enhanced pickup ion-induced depletion on the nightside during higher dynamic pressure for CIR event. The higher dynamic pressure during specific CIR event leads to more electron and ion density depletions in a greater number of orbit profiles in the current study (Figures 6 & 7). The nightside ionosphere depends upon the electron impact ionization and/or day-to-night transport processes to sustain plasma at higher altitudes (Cao et al., 2019; Cui et al., 2015; Girazian et al., 2017). The compression in the dayside leads to the inhibition of plasma transport from day to nightside (Cravens et al.,1982; Haider, 1997; Miller et al., 1984). This situation results in the lower ionopause altitude on the nightside during specific CME and CIR events for our study.

The primary aim of the present study is to compare the Martian ionospheric response to incoming CMEs and CIRs during the declining phase of the Solar cycle 24 (2015-2020). We have extensively analyzed the ionospheric variation in both dayside and nightside during 15

CMEs and 15 CIRs. The notable difference in the mean ion density profiles during dayside and nightside is due to the different production mechanisms for both sides of the ionosphere (Figures 8 & 9). The mean ion density profiles during both CMEs and CIRs show a significant deviation from the mean quiet-time profiles. In addition, our analysis (for 15 CMEs and 15 CIRs) clearly indicate that Martian ionospheric depletion is more prominent during CIRs in comparison to CMEs (Figures 8 & 9). The difference between mean density profiles (shown in red and blue colors) during CMEs and CIRs is less pronounced on the dayside as compared to the nightside. The lesser difference in the mean density profiles during the dayside might be due to consistent ion production by photoionization and ion-neutral chemistry during the daytime (Haider & Mahajan, 2014; Martinis et al., 2003; Singh & Prasad, 1983). During the dayside and nightside, there is a difference in the behavior of the heavier ($O_2^+$, $CO_2^+$, & $NO^+$) and lighter ions ($O^+$, $C^+$, $N^+$, & $OH^+$). For the heavier ion, the density profiles show a sharp decrease above 250 km altitude for both CMEs and CIRs. Whereas, lighter ions show less steep density profiles with respect to altitude. The mean peak altitude for all lighter ions ($O^+$, $C^+$, $N^+$, & $OH^+$) existed at a lower altitude during CIRs as compared to CMEs. The comparative behavior of the lighter ions ($C^+$, $N^+$, & $OH^+$) is studied for the first time to the best of our knowledge in the dayside and nightside ionosphere during CMEs and CIRs. The mean quiet-time peak altitude and density (275 km and 1079.69 $cm^{-3}$) were more in comparison to both CMEs and CIRs. Further, during CIRs, the $O^+$ dayside mean peak altitude and density (240 km and 844.52 $cm^{-3}$) were less in comparison to the CMEs (265 km and 1017.03 $cm^{-3}$). It has been reported that during the moderate solar cycle, the peak density of $O^+$ lies nearly at 293-300 km (Fox et al., 2021; Withers et al., 2015). Interestingly, it occurred at a lower altitude in the present case. During both CMEs and CIRs, the mean peak altitude of lighter ions for the nightside was always lower compared to the dayside. The peak density of the heavier ions ($O_2^+$, $CO_2^+$, & $NO^+$) was mostly at around 110-130 km as reported by Bougher et al. (2015). Since MAVEN measured the ionospheric plasma mostly between 150-500 km altitude range, we are unable to comment on the mean peak altitude with peak density of the heavier ions. The above discussion indicates the more impact of CIRs on Martian ionospheric species compared to CMEs. As there are significant differences between two major solar events i.e., CME and CIR. The strength of the magnetospheric convection electric field of CME events is often stronger than that of CIRs (Borovsky & Denton, 2006). However, the duration of CIRs is usually much longer than CMEs. The studies of Turner et al. (2009) and Emery et al. (2009) have shown that these less-intense and long-duration CIR events thus can deposit roughly the same amount of energy or even more energy into the upper atmosphere than most of the moderate CMEs do over the entire events on the Earth's Magnetosphere-ionosphere system. Edberg et al. (2010) observed that during CIRs at Mars, the plasma outflow increases by a factor of 2.5 and increases the loss rate by 22.5%.

It is known that the solar wind is an important driver that greatly affects the Martian upper ionosphere. The topside ionosphere gets more depleted due to high solar wind dynamic pressure at all solar zenith angles. The previous studies also suggest that high solar wind dynamic pressures lead to ionospheric compression, increased ion escape, and reduced day-to-night plasma transport to the high-altitude nightside ionosphere (Dubinin et al., 2018; Girazian et al., 2019; Sánchez-Cano et al., 2020). In addition to the solar wind dynamic pressure, IMF can also play an important role in the variation of the Martian ionosphere. In previous studies, it has been reported that during CIRs, when the solar wind dynamic pressure is high, the IMF penetrates deep into the dayside ionosphere and molecular ion escape can be observed (Dubinin et al., 2009; Girazian et al., 2019; Kubota et al., 2013). During many CIRs, the magnetic lines drape around the planet in the pileup region of Mars, and change the orientation such that the oppositely directed magnetic field lines can verge upon with each other (Cravens et al., 2020; Edberg et al., 2010). This leads to a reconnection related escape mechanism. Also, reconnection mechanism is more efficient above 300 km (collision-less processes dominated) and below 300 km collisional processes reduce the efficiency of reconnection. Further, while draping of the magnetic field around Mars, $\mathbf{J} \times \mathbf{B}$ forces could increase the plasma erosion channels (Dubinin et al., 2009; Edberg et al., 2010). This accelerates the plasma and hence could be one of the factors to enhance the escape rate. Therefore, the dynamic pressure, reconnection related escape mechanism and $\mathbf{J} \times \mathbf{B}$ forces at higher altitudes play an important role to accelerate and energize the plasma in the Martian ionosphere (Ergun et al., 2006; Fowler et al., 2019; Girazian et al., 2017, 2019; Opgenoorth et al., 2013; Sánchez-Cano et al., 2020). Furthermore, it would consequently lead to the lower plasma density and attain one order less magnitude in the topside ionosphere during CIRs in comparison to CMEs in our study. There are other important factors like low-frequency plasma waves and ion pick up which also contribute to drive the loss processes during both events (Ergun et al., 2006; Lundin et al., 2008). This possibility could be investigated in the future for a better understanding of the loss processes in the Martian ionosphere. Based on the present study, we observed that the CMEs and CIRs have a significant impact on the Martian ionospheric species during the declining phase of solar cycle 24. The impact during CIRs is more prominent in comparison to the CMEs. Further, the depletion is more pronounced in the case of the nightside compared to the dayside. However, further investigation is needed to understand why the CIR causes more impact than the CME on the Martian ionosphere although the CMEs are known to be more intense events. This will be taken up for future investigation.

## 5. Summary and conclusions

The impact of CMEs and CIRs on the Martian ionosphere during the declining (2015-2020) phase of Solar cycle 24 is investigated using the observations from the MAVEN datasets. The CME and CIR events were observed at Mars with enhancements in the solar wind density, velocity, solar wind dynamic pressure, and fluctuating IMF. From our observations, we found that the specific CME (03-05 March 2015) and CIR (19-21 March 2018) cause more depletion during the nightside as compared to the dayside. The sharp density gradient in the plasma indicates the formation of ionopause for orbits 822, 826 (dayside, CME), 823 (nightside, CME), 6748 & 6749 (dayside, CIR), and 6750 & 6751 (nightside, CIR). Due to higher peak dynamic pressure during disturb solar wind conditions, the depletion in the plasma density occurs for orbit profiles during specific CME and CIR events. In addition, the primary focus of the paper is to provide a comparative average scenario of the Martian ionospheric response during 15 CMEs and 15 CIRs. The average analyses indicate that the ionospheric species during CIRs are more depleted compared to CMEs. During CIRs, nightside mean plasma density profiles are more depleted and of one order less magnitude (above 250 km) than during CMEs. The mean peak altitude of lighter ions ($O^+$, $C^+$, $N^+$, & $OH^+$) is lesser during CIRs and CMEs as compared to mean quiet-time. Further, the peak mean altitude and density for lighter ions are lesser on the dayside and nightside during CIRs in comparison to CMEs. In addition, the depletion is more pronounced in the case of the nightside with lower mean peak altitude and density compared to the dayside. The dynamic pressure, IMF, and solar wind convection electric field persist for a longer time for CIRs than CMEs which may lead to more depletion of the Martian upper ionosphere. Therefore, this study indicates that the impact of CIRs on the Martian ionospheric species is more prominent compared to CMEs.

## Data availability statement

We have used the NASA DONKI catalog for space weather forecast information. The MAVEN in-situ key parameter datasets: The LPW (electron density and temperature) Level 2, version_02, revision_01 (Andersson, 2017), NGIMS (ions density) Level 2, version_08, revision_01 (Benna & Lyness, 2014), SWIA (solar wind ion) Level 2, version_01, revision_01 (Halekas, 2017), and MAG (IMF) Level 2,

version_01, revision_01 (Connerney, 2017) datasets used in this work are taken from the NASA Planetary Data System (PDS) and can be downloaded using the Python Data Analysis and Visualization tool (MAVEN SDC et al., 2020). The WSA-ENLIL+Cone model simulation and derived data products used and produced during this work can be found in Ram et al. (2023).

**Acknowledgements**

We sincerely acknowledge the MAVEN team for the data. We also acknowledge the WSA-ENLIL+CONE Model for simulation snapshots. L. Ram acknowledges the fellowship from the Ministry of Education, Government of India for carrying out this research work. D. Rout acknowledges the support from Humboldt Research Fellowship for Postdoctoral Researchers (Humboldt foundation grants PSP D-023-20- 001). R. Rathi acknowledges the fellowship from the Innovation in Science Pursuit for Inspired Research (INSPIRE) programme, Department of Science and Technology, Government of India. This work is also supported by the Ministry of Education, Government of India.

## References:

Andersson, L., Ergun, R. E., Delory, G. T., Eriksson, A., Westfall, J., Reed, H., et al. (2015). The Langmuir Probe and Waves (LPW) Instrument for MAVEN. *Space Science Reviews*, *195*(1-4), 173–198. Doi:10.1007/s11214-015-0194-3

Andersson, L. (2017). MAVEN Langmuir Probe and Waves (LPW) Bundle. NASA Planetary Data System. [Dataset]. https://doi.org/10.17189/1410658

Belcher, J. W., & Davis Jr, L. (1971). Large‐amplitude Alfvén waves in the interplanetary medium, 2. *Journal of Geophysical Research*, *76*(16), 3534-3563. https://doi.org/10.1029/JA076i016p03534

Benna, M., Lyness, E. (2014), MAVEN Neutral Gas and Ion Mass Spectrometer Data, NASA Planetary Data System. [Dataset]. https://doi.org/10.17189/1518931

Benna, M., Mahaffy, P. R., Grebowsky, J. M., Fox, J. L., Yelle, R. V., & Jakosky, B. M. (2015). First measurements of composition and dynamics of the Martian ionosphere by MAVEN's Neutral Gas and Ion Mass Spectrometer. *Geophysical Research Letters*, *42*(21), 8958-8965. https://doi.org/10.1002/2015GL066146

Borovsky, J. E., & Denton, M. H. (2006). Differences between CME‐driven storms and CIR‐driven storms. *Journal of Geophysical Research: Space Physics*, *111*(A7). https://doi.org/10.1029/2005JA011447

Bougher, S. W., Pawlowski, D., Bell, J. M., Nelli, S., McDunn, T., Murphy, J. R., et al. (2015). Mars Global Ionosphere‐Thermosphere Model: Solar cycle, seasonal, and diurnal variations of the Mars upper atmosphere. *Journal of Geophysical Research: Planets*, *120*(2), 311-342. https://doi.org/10.1002/2014JE004715

Burlaga, L. F. (1974). Interplanetary stream interfaces. *Journal of Geophysical Research*, *79*(25), 3717-3725. https://doi.org/10.1029/JA079i025p03717

Cane, H. V., & Richardson, I. G. (2003). Interplanetary coronal mass ejections in the near‐Earth solar wind during 1996–2002. *Journal of Geophysical Research: Space Physics*, *108*(A4). https://doi.org/10.1029/2002JA009817

Cao, Y. T., Cui, J., Wu, X. S., Guo, J. P., & Wei, Y. (2019). Structural variability of the nightside Martian ionosphere near the terminator: Implications on plasma sources. *Journal of Geophysical Research: Planets, 124*(6), 1495–1511. https://doi.org/10.1029/2019JE005970

Chen, P. F. (2011). Coronal mass ejections: models and their observational basis. *Living Reviews in Solar Physics*, *8*(1), 1-92. https://doi.org/10.12942/lrsp-2011-1

Chu, F., Girazian, Z., Duru, F., Ramstad, R., Halekas, J., Gurnett, D. A., et al. (2021). The dayside ionopause of mars: Solar wind interaction, pressure balance, and comparisons with Venus. *Journal of Geophysical Research: Planets*, *126*(11). https://doi.org/10.1029/2021JE006936

Cloutier, P. A., McElroy, M. B., & Michel, F. C. (1969). Modification of the Martian ionosphere by the solar wind. *Journal of Geophysical Research*, *74*(26), 6215-6228. https://doi.org/10.1029/JA074i026p06215

Connerney, J. E., Espley, J. R., DiBraccio, G. A., Gruesbeck, J. R., Oliversen, R. J., Mitchell, D. L., et al. (2015). First results of the MAVEN magnetic field investigation. *Geophysical Research Letters*, *42*(21), 8819-8827. https://doi.org/10.1002/2015GL065366

Connerney, J. E., Espley, J., Lawton, P., Murphy, S., Odom, J., Oliversen, R., et al. (2015). The MAVEN magnetic field investigation. *Space Science Reviews*, *195*(1), 257-291. https://doi.org/10.1007/s11214-015-0169-4

Connerney, J. E. (2017). MAVEN Magnetometer (MAG) Calibrated Data Bundle. NASA Planetary Data System. [Dataset]. https://doi.org/10.17189/1414178

Cravens, T. E., L. H. Brace, H. A. Taylor Jr, C. T. Russell, W. L. Knudsen, K. L. Miller, et al. (1982). Disappearing ionospheres on the nightside of Venus. *Icarus* 51, no. 2, 271-282. https://doi.org/10.1016/0019-1035(82)90083-5

Cravens, T. E., Fowler, C. M., Brain, D., Rahmati, A., Xu, S., Ledvina, S. A., et al. (2020). Magnetic reconnection in the ionosphere of Mars: The role of collisions. *Journal of Geophysical Research: Space Physics*, *125*(9). https://doi.org/10.1029/2020JA028036

Cui, J., Galand, M., Yelle, R. V., Wei, Y., & Zhang, S. J. (2015). Day-to-night transport in the Martian ionosphere: Implications from total electron content measurements. *Journal of Geophysical Research: Space Physics*, *120*(3), 2333-2346. https://doi.org/10.1002/2014JA020788

Dubinin, E., Fraenz, M., Woch, J., Duru, F., Gurnett, D., Modolo, R., et al. (2009). Ionospheric storms on Mars: Impact of the corotating interaction region. *Geophysical Research Letters*, *36*(1). https://doi.org/10.1029/2008GL036559

Dubinin, E. M., Sauer, K., Baumgärtel, K., & Lundin, R. (1997). The Martian magnetosheath: PHOBOS-2 observations. *Advances in Space Research*, *20*(2), 149-153. https://doi.org/10.1016/S0273-1177(97)00525-5

Dubinin, E., Fränz, M., Pätzold, M., McFadden, J., Halekas, J. S., Connerney, J. E. P., et al. (2018). Martian ionosphere observed by MAVEN. 3. Influence of solar wind and IMF on upper ionosphere. *Planetary and Space Science*, *160*, 56-65. https://doi.org/10.1016/j.pss.2018.03.016

Duru, F., Baker, N., De Boer, M., Chamberlain, A., Verchimak, R., Morgan, D. D., et al. (2020). Martian ionopause boundary: Coincidence with photoelectron boundary and response to internal and external drivers. *Journal of Geophysical Research: Space Physics*, *125*(5), e2019JA027409

Edberg, N. J. T., Nilsson, H., Williams, A. O., Lester, M., Milan, S. E., Cowley, S. W. H., et al. (2010). Pumping out the atmosphere of Mars through solar wind pressure pulses. *Geophysical Research Letters*, *37*(3). https://doi.org/10.1029/2009GL041814

Emery, B. A., Richardson, I. G., Evans, D. S., & Rich, F. J. (2009). Solar wind structure sources and periodicities of auroral electron power over three solar cycles. *Journal of Atmospheric and Solar-Terrestrial Physics*, *71*(10-11), 1157-1175. doi:10.1016/j.jastp.2008.08.005

ENLIL Solar Wind Prediction. (2010, March). [Software]. http://helioweather.net/

Ergun, R. E., Andersson, L., Peterson, W. K., Brain, D., Delory, G. T., Mitchell, D. L., et al. (2006). Role of plasma waves in Mars' atmospheric loss. *Geophysical research letters*, *33*(14). https://doi.org/10.1029/2006GL025785

Fowler, C. M., Lee, C. O., Xu, S., Mitchell, D. L., Lillis, R., Weber, T., et al. (2019). The penetration of draped magnetic field into the Martian upper ionosphere and correlations with upstream solar wind dynamic pressure. *Journal of Geophysical Research: Space Physics*, *124*(4), 3021-3035. https://doi.org/10.1029/2019JA026550

Fox, J. L., & Dalgarno, A. (1979). Ionization, luminosity, and heating of the upper atmosphere of Mars. *Journal of Geophysical Research: Space Physics*, *84*(A12), 7315-7333. https://doi.org/10.1029/JA084iA12p07315

Fox, J. L., Benna, M., McFadden, J. P., & Jakosky, B. M. (2021). Rate coefficients for the reactions of $CO_2^+$ with O: Lessons from MAVEN at Mars. *Icarus*, *358*, 114186. https://doi.org/10.1016/j.icarus.2020.114186

Geyer P., Temmer M., Guo J., & Heinemann, S. (2021), Properties of stream interaction regions at Earth and Mars during the declining phase of SC 24, *A&A*, *649* (A80), 20. https://doi.org/10.1051/0004-6361/202040162

Girazian, Z., Mahaffy, P., Lillis, R. J., Benna, M., Elrod, M., Fowler, C. M., et al. (2017). Ion densities in the nightside ionosphere of Mars: Effects of electron impact ionization. *Geophysical Research Letters*, *44*(22), 11-248

Girazian, Z., Halekas, J., Morgan, D. D., Kopf, A. J., Gurnett, D. A., & Chu, F. (2019). The effects of solar wind dynamic pressure on the structure of the topside ionosphere of Mars. *Geophysical Research Letters*, *46*(15), 8652-8662. https://doi.org/10.1029/2019GL083643

Gopalswamy, N. (2006). Coronal mass ejections of solar cycle 23. *Journal of Astrophysics and Astronomy*, *27*(2), 243-254. https://doi.org/10.1007/BF02702527

Gosling, J. T. (1990). Coronal mass ejections and magnetic flux ropes in interplanetary space. *Physics of magnetic flux ropes*, *58*, 343-364. https://doi.org/10.1029/GM058p0343

Gosling, J. T., & Forsyth, R. J. (2001). CME-driven solar wind disturbances at high heliographic latitudes. *The 3-D Heliosphere at Solar Maximum*, 87-98. https://doi.org/10.1007/978-94-017-3230-7_15

Gosling, J. T., McComas, D. J., Phillips, J. L., & Bame, S. J. (1991). Geomagnetic activity associated with Earth passage of interplanetary shock disturbances and coronal mass ejections. *Journal of Geophysical Research: Space Physics*, *96*(A5), 7831-7839. https://doi.org/10.1029/91JA00316

Gosling, J. T., & Pizzo, V. J. (1999). Formation and evolution of corotating interaction regions and their three dimensional structure. *Corotating interaction regions*, 21-52. https://doi.org/10.1007/978-94-017-1179-1_3

Haider, S. A. (1997). Chemistry of the nightside ionosphere of Mars. *Journal of Geophysical Research: Space Physics*, *102*(A1), 407-416. https://doi.org/10.1029/96JA02353

Haider, S. A., & Mahajan, K. K. (2014). Lower and upper ionosphere of Mars. *Space Science Reviews*, *182*(1), 19-84. https://doi.org/10.1007/s11214-014-0058-2

Halekas, J. S., Taylor, E. R., Dalton, G., Johnson, G., Curtis, D. W., McFadden, J. P., et al. (2015). The solar wind ion analyzer for MAVEN. *Space Science Reviews*, *195*(1), 125-151. https://doi.org/10.1007/s11214-013-0029-z

Halekas, J. S., Ruhunusiri, S., Harada, Y., Collinson, G., Mitchell, D. L., Mazelle, C., et al. (2017). Structure, dynamics, and seasonal variability of the Mars-solar wind interaction: MAVEN Solar Wind Ion Analyzer in-flight performance and science results. *Journal of Geophysical Research: Space Physics*, *122*(1), 547-578. https://doi.org/10.1002/2016JA023167

Halekas, J. (2017). MAVEN Solar Wind Ion Analyzer (SWIA) Calibrated Data Bundle. NASA Planetary Data System. [Dataset]. https://doi.org/10.17189/1414182

Hara, T., Seki, K., Futaana, Y., Yamauchi, M., Yagi, M., Matsumoto, Y., et al. (2011). Heavy-ion flux enhancement in the vicinity of the Martian ionosphere during CIR passage: Mars Express ASPERA-3 observations. *Journal of Geophysical Research: Space Physics*, *116*(A2). https://doi.org/10.1029/2010JA015778

Huang, H., Guo, J., Wang, Z., Lin, H., Zheng, J., Cui, J., et al. (2019). Properties of stream interactions and their associated shocks near 1.52 au: MAVEN Observations. *The Astrophysical Journal*, *879*(2), 118. https://doi.org/10.3847/1538-4357/ab25e9

Jakosky, B. M., Grebowsky, J. M., Luhmann, J. G., Connerney, J., Eparvier, F., Ergun, R., et al. (2015). MAVEN observations of the response of Mars to an interplanetary coronal mass ejection. *Science*, *350*(6261). DOI: 10.1126/science.aad0210

Jethwa M., (2021). Development, Implementation and Testing of Algorithm to Detect Ionopause-Like Structure Using Parallel Processing in the Martian Atmosphere. *IJRASET, 9*(XII). https://doi.org/10.22214/ijraset.2021.39653

Jian, L. C. T. R., Russell, C. T., Luhmann, J. G., & Skoug, R. M. (2006). Properties of interplanetary coronal mass ejections at one AU during 1995–2004. *Solar Physics*, *239*(1), 393-436. https://doi.org/10.1007/s11207-006-0133-2

Krishnaprasad, C., Thampi, S. V., & Bhardwaj, A. (2019). On the response of Martian ionosphere to the passage of a corotating interaction region: MAVEN observations. *Journal of Geophysical Research: Space Physics*, *124*(8), 6998-7012. https://doi.org/10.1029/2019JA026750

Krishnaprasad, C., Thampi, S. V., Bhardwaj, A., Pant, T. K., & Thampi, R. S. (2021). Ionospheric plasma energization at Mars during the September 2017 ICME event. *Planetary and Space Science*, *205*, 105291. https://doi.org/10.1016/j.pss.2021.105291

Kubota, Y., Maezawa, K., Jin, H., & Fujimoto, M. (2013). IMF-induced escape of molecular ions from the Martian ionosphere. In *Annales Geophysical* (Vol. 31, No. 8, pp. 1343-1356). Copernicus GmbH. https://doi.org/10.5194/angeo-31-1343-2013

Lee, C. O., Hara, T., Halekas, J. S., Thiemann, E., Chamberlin, P., Eparvier, F., & Jakosky, B. M. (2017). MAVEN observations of the solar cycle 24 space weather conditions at Mars. *Journal of Geophysical Research: Space Physics*, *122*(3), 2768-2794. https://doi.org/10.1002/2016JA023495

Lundin, R., Zakharov, A., Pellinen, R., Barabasj, S. W., Borg, H., Dubinin, E. M., et al. (1990). ASPERA/Phobos measurements of the ion outflow from the Martian ionosphere. *Geophysical research letters*, *17*(6), 873-876. https://doi.org/10.1029/GL017i006p00873

Lundin, R., Barabash, S., Holmstrom, M., Nilsson, H., Yamauchi, M., Fraenz, M., et al. (2008). A comet-like escape of ionospheric plasma from Mars. *Geophysical Research Letters*, *35*(18). https://doi.org/10.1029/2008GL034811

Ma, Y. J., & Nagy, A. F. (2007). Ion escape fluxes from Mars. *Geophysical Research Letters*, *34*(8). https://doi.org/10.1029/2006GL029208

Mahaffy, P. R., Benna, M., King, T., Harpold, D. N., Arvey, R., Barciniak, M., et al. (2015). The neutral gas and ion mass spectrometer on the Mars atmosphere and volatile evolution mission. *Space Science Reviews*, *195*(1), 49-73. https://doi.org/10.1007/s11214-014-0091-1

Martinis, C. R., Wilson, J. K., & Mendillo, M. J. (2003). Modeling day-to-day ionospheric variability on Mars. *Journal of Geophysical Research: Space Physics*, *108*(A10). https://doi.org/10.1029/2003JA009973

MAVEN SDC, Harter, B., Ben, Elysia Lucas, & Jibarnum. (2020). MAVENSDC/Pydivide: First Release (Version v0.2.13). [Software]. Zenodo. https://doi.org/10.5281/ZENODO.3601516

Mays, M. L., Thompson, B. J., Jian, L. K., Colaninno, R. C., Odstrcil, D., Möstl, C., et al. (2015). Propagation of the 2014 January 7 CME and resulting geomagnetic non-event. *The Astrophysical Journal*, *812*(2), 145. Doi: 10.1088/0004-637X/812/2/145

McElroy, M. B., Kong, T. Y., & Yung, Y. L. (1977). Photochemistry and evolution of Mars' atmosphere: A Viking perspective. *Journal of Geophysical Research*, *82*(28), 4379-4388. https://doi.org/10.1029/JS082i028p04379

Michel, F. C. (1971). Solar wind interaction with planetary atmospheres. *Reviews of Geophysics*, *9*(2), 427-435. https://doi.org/10.1029/RG009i002p00427

Miller, K. L., Knudsen, W. C., & Spenner, K. (1984). The dayside Venus ionosphere: I. Pioneer-Venus retarding potential analyzer experimental observations. *Icarus*, *57*(3), 386-409. https://doi.org/10.1016/0019-1035(84)90125-8

Morgan, D. D., Gurnett, D. A., Kirchner, D. L., Winningham, J. D., Frahm, R. A., Brain, D. A., et al. (2010). Radar absorption due to a corotating interaction region encounter with Mars detected by MARSIS. *Icarus*, *206*(1), 95-103. https://doi.org/10.1016/j.icarus.2009.03.008

Neugebauer, M., Goldstein, R., & Goldstein, B. E. (1997). Features observed in the trailing regions of interplanetary clouds from coronal mass ejections. *Journal of Geophysical Research: Space Physics*, *102*(A9), 19743-19751. https://doi.org/10.1029/97JA01651

Niu, D., Gu, H., Cui, J., Wu, X., Wu, M., & Zhang, T. (2021). Effects of the Solar Wind Dynamic Pressure on the Martian Topside Ion Distribution: Implications on the Variability of Bulk Ion Outflow. *The Astrophysical Journal*, *922*(2), 231. https://doi.org/10.3847/1538-4357/ac2bfc

Odstrcil, D. (2003). Modeling 3-D solar wind structure. *Advances in Space Research*, *32*(4), 497-506. https://doi.org/10.1016/S0273-1177(03)00332-6

Opgenoorth, H. J., Andrews, D. J., Fränz, M., Lester, M., Edberg, N. J. T., Morgan, D., et al. (2013). Mars ionospheric response to solar wind variability. *Journal of Geophysical Research: Space Physics*, *118*(10), 6558-6587. https://doi.org/10.1002/jgra.50537

Pizzo, V. (1978). A three-dimensional model of corotating streams in the solar wind 1. Theoretical foundations *Journal of Geophysical Research*, 83, 5563-5572. https://doi.org/10.1029/JA083iA12p05563

Ram, L., Rout, D., Rathi, R., Mondal, S., Sarkhel, S., & Halekas, J. (2023). A comparison of the impacts of CMEs and CIRs on the Martian dayside and nightside ionospheric species. [Dataset]. Zenodo. https://doi.org/10.5281/zenodo.7693250

Richardson, I. G. (2018). Solar wind stream interaction regions throughout the heliosphere. *Living reviews in solar physics*, *15*(1), 1-95. https://doi.org/10.1007/s41116-017-0011-z

Rout, D., Chakrabarty, D., Janardhan, P., Sekar, R., Maniya, V., & Pandey, K. (2017). Solar wind flow angle and geoeffectiveness of corotating interaction regions: First results. *Geophysical Research Letters, 44*(10), 4532-4539. https://doi.org/10.1002/2017GL073038

Russell, C. T., Shinde, A. A., & Jian, L. (2005). A new parameter to define interplanetary coronal mass ejections. *Advances in Space Research, 35*(12), 2178-2184. https://doi.org/10.1016/j.asr.2005.04.024

Sánchez-Cano, B., Narvaez, C., Lester, M., Mendillo, M., Mayyasi, M., Holmstrom, M., et al. (2020). Mars ionopause: A matter of pressures. *Journal of Geophysical Research: Space Physics, 125*(9). https://doi.org/10.1029/2020JA028145

Schunk, R., & Nagy, A. (2009). *Ionospheres: physics, plasma physics, and chemistry*. Cambridge university press. www.cambridge.org/9780521877060

Singh, R. N., & Prasad, R. (1983). Production of dayside ionosphere of Mars. *Journal of Astrophysics and Astronomy, 4*(4), 261-269. https://doi.org/10.1007/BF02714920

Smith, E. J., & Wolfe, J. H. (1976). Observations of interaction regions and corotating shocks between one and five AU: Pioneers 10 and 11. *Geophysical Research Letters, 3*(3), 137-140. https://doi.org/10.1029/GL003i003p00137

Space Weather Database of Notifications, Knowledge, and Information (DONKI). [Software]. https://kauai.ccmc.gsfc.nasa.gov/DONKI/search/

Thampi, S. V., Krishnaprasad, C., Bhardwaj, A., Lee, Y., Choudhary, R. K., & Pant, T. K. (2018). MAVEN Observations of the response of Martian ionosphere to the interplanetary coronal mass ejections of March 2015. *Journal of Geophysical Research: Space Physics, 123*(8), 6917-6929. https://doi.org/10.1029/2018JA025444

Thampi, S. V., Krishnaprasad, C., Nampoothiri, G. G., & Pant, T. K. (2021). The impact of a stealth CME on the Martian topside ionosphere. *Monthly Notices of the Royal Astronomical Society, 503*(1), 625-632. https://doi.org/10.1093/mnras/stab494

Tousey, R. (1973). The solar corona. In *Space Research Conference* (Vol. 2, pp. 713-730).

Turner, N. E., Cramer, W. D., Earles, S. K., & Emery, B. A. (2009). Geoefficiency and energy partitioning in CIR-driven and CME-driven storms. *Journal of Atmospheric and Solar-Terrestrial Physics, 71*(10-11), 1023-1031. https://doi.org/10.1016/j.jastp.2009.02.005

Vogt, M. F., Withers, P., Mahaffy, P. R., Benna, M., Elrod, M. K., Halekas, J. S., et al. (2015). Ionopause-like density gradients in the Martian ionosphere: A first look with MAVEN. *Geophysical Research Letters, 42*(21), 8885-8893. https://doi.org/10.1002/2015GL065269

Wei, Y., Fraenz, M., Dubinin, E., Woch, J., Lühr, H., Wan, W., et al. (2012). Enhanced atmospheric oxygen outflow on Earth and Mars driven by a corotating interaction region. *Journal of Geophysical Research: Space Physics, 117*(A3). https://doi.org/10.1029/2011JA017340

Wimmer-Schweingruber, R. F., Crooker, N. U., Balogh, A., Bothmer, V., Forsyth, R. J., Gazis, P., et al. (2006). Understanding interplanetary coronal mass ejection signatures. *Coronal Mass Ejections*, 177-216. https://doi.org/10.1007/978-0-387-45088-9_10

Withers, P., Vogt, M., Mahaffy, P., Benna, M., Elrod, M., & Jakosky, B. M. (2015). Changes in the thermosphere and ionosphere of Mars from Viking to MAVEN. *Geophysical Research Letters, 42*(21), 9071-9079. https://doi.org/10.1002/2015GL065985

Withers, P., Felici, M., Mendillo, M., Moore, L., Narvaez, C., Vogt, M. F., et al. (2018). First ionospheric results from the MAVEN radio occultation science experiment (ROSE). *Journal of Geophysical Research: Space Physics*, *123*(5), 4171-4180. https://doi.org/10.1029/2018JA025182

Zhao, D., Guo, J., Huang, H., Lin, H., Hong, Y., Feng, X., et al. (2021). Interplanetary Coronal Mass Ejections from MAVEN Orbital Observations at Mars. *The Astrophysical Journal*, *923*(1), 4. https://doi.org/10.3847/1538-4357/ac294b

Zurbuchen, T. H., & Richardson, I. G. (2006). In-situ solar wind and magnetic field signatures of interplanetary coronal mass ejections. *Coronal mass ejections*, 31-43. https://doi.org/10.1007/978-0-387-45088-9_3

**Figures:**

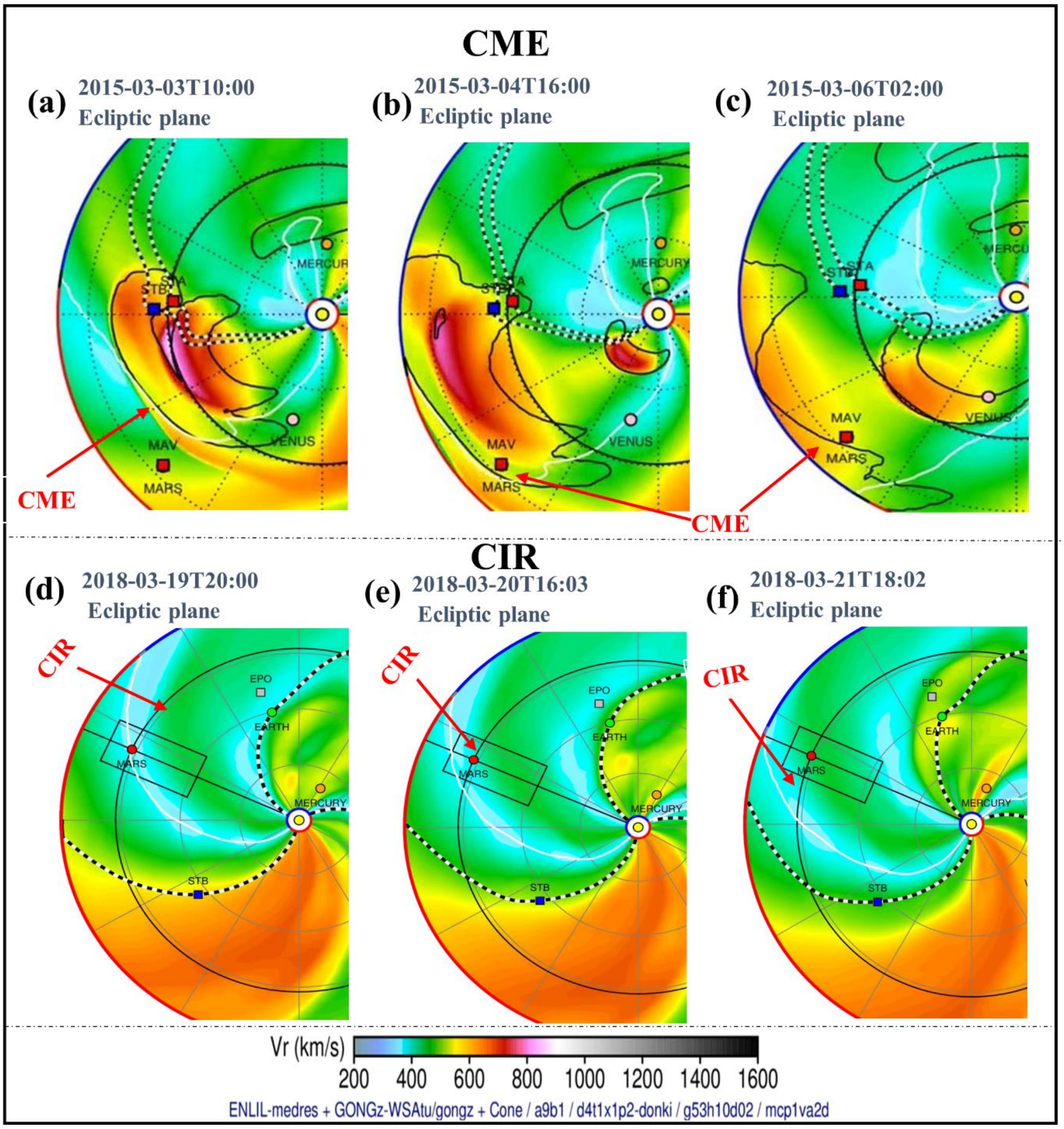

**Figure 1:** The WSA-ENLIL+ Cone model inner heliospheric simulation screenshots, (a) before the event (b) during the event, and (c) after the event, showing the solar wind radial velocity (color contour) and interplanetary magnetic field (IMF) during the 03-05 March 2015 CME event (pointed by red arrow) and (d) before the event, (e) during the event, (f) after the event, showing the solar wind radial velocity (color contour) and interplanetary magnetic field (IMF) during the 19-21 March 2018 CIR event (pointed by red arrow). IMF lines are shown by the black–white zebra lines. A CME (top panel) is outlined by the black contour and the heliospheric current sheet (HCS) is shown by the white line.

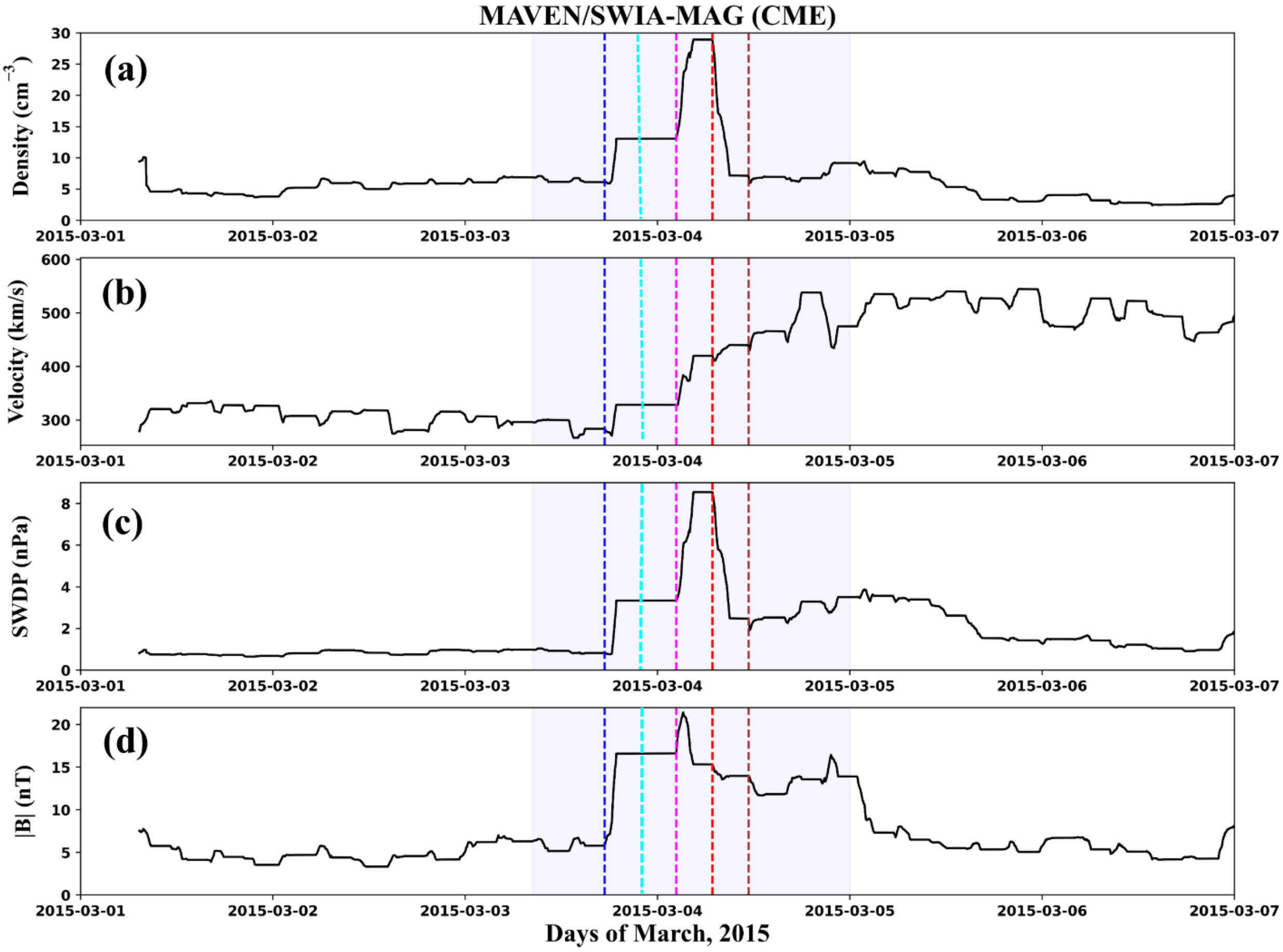

**Figure 2:** (a) Solar wind density, (b) solar wind velocity, (c) dynamic pressure, and (d) IMF(|B|), observations near Mars, during the 03-05 March 2015 CME event. The vertical-colored dotted lines from left to right in each panel indicate the MAVEN orbits 822 (blue), 823 (cyan), 824 (magenta), 825 (red), and 826 (brown) during disturb time.

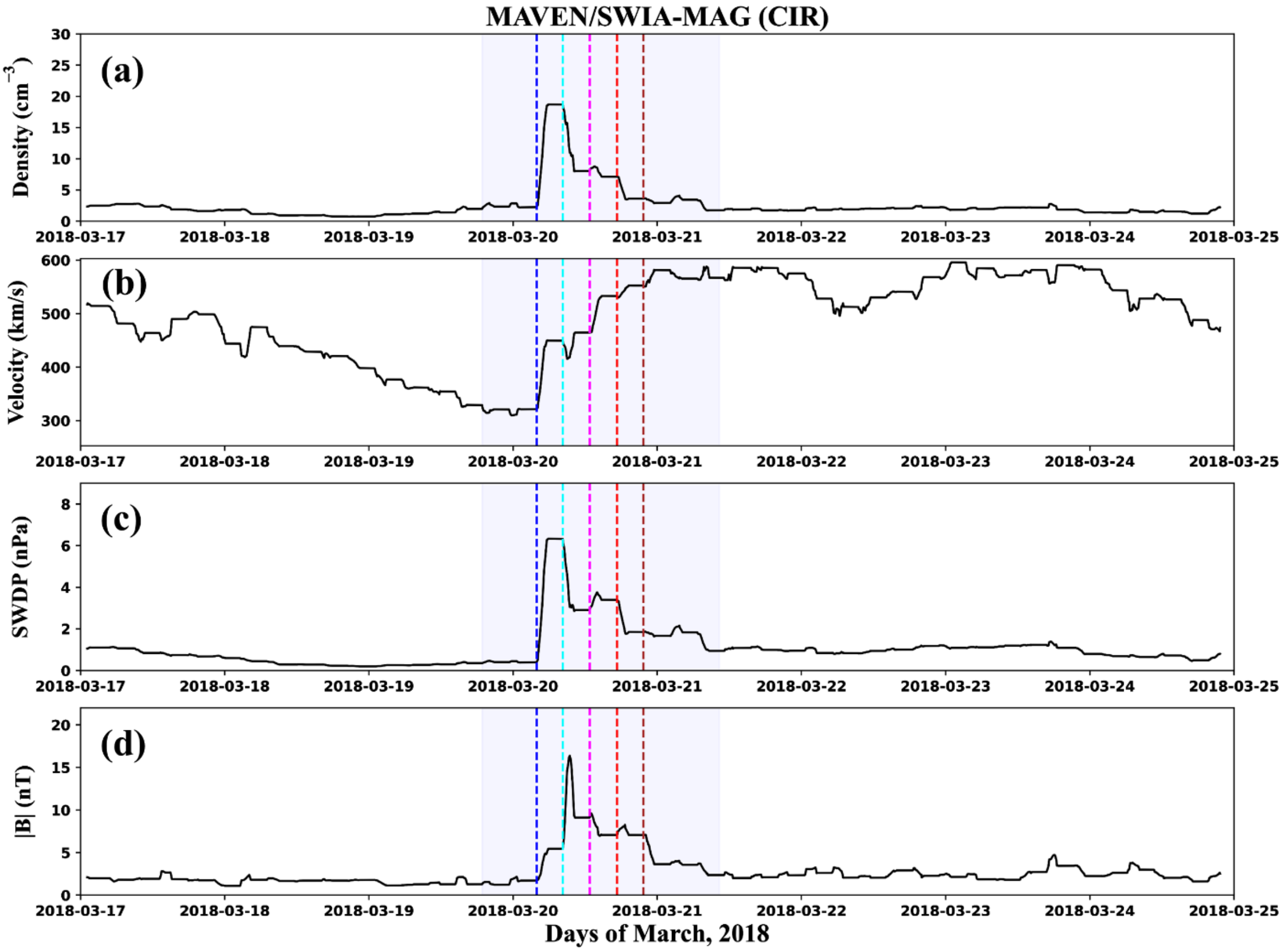

**Figure 3:** (a) Solar wind density, (b) solar wind velocity, (c) dynamic pressure, and (d) IMF(|B|), observations near Mars, during the 19-21 March 2018 CIR event. The vertical-colored dotted lines from left to right in each panel indicate the MAVEN orbits 6748 (blue), 6749 (cyan), 6750 (magenta), 6751 (red), and 6752 (brown) during disturb time.

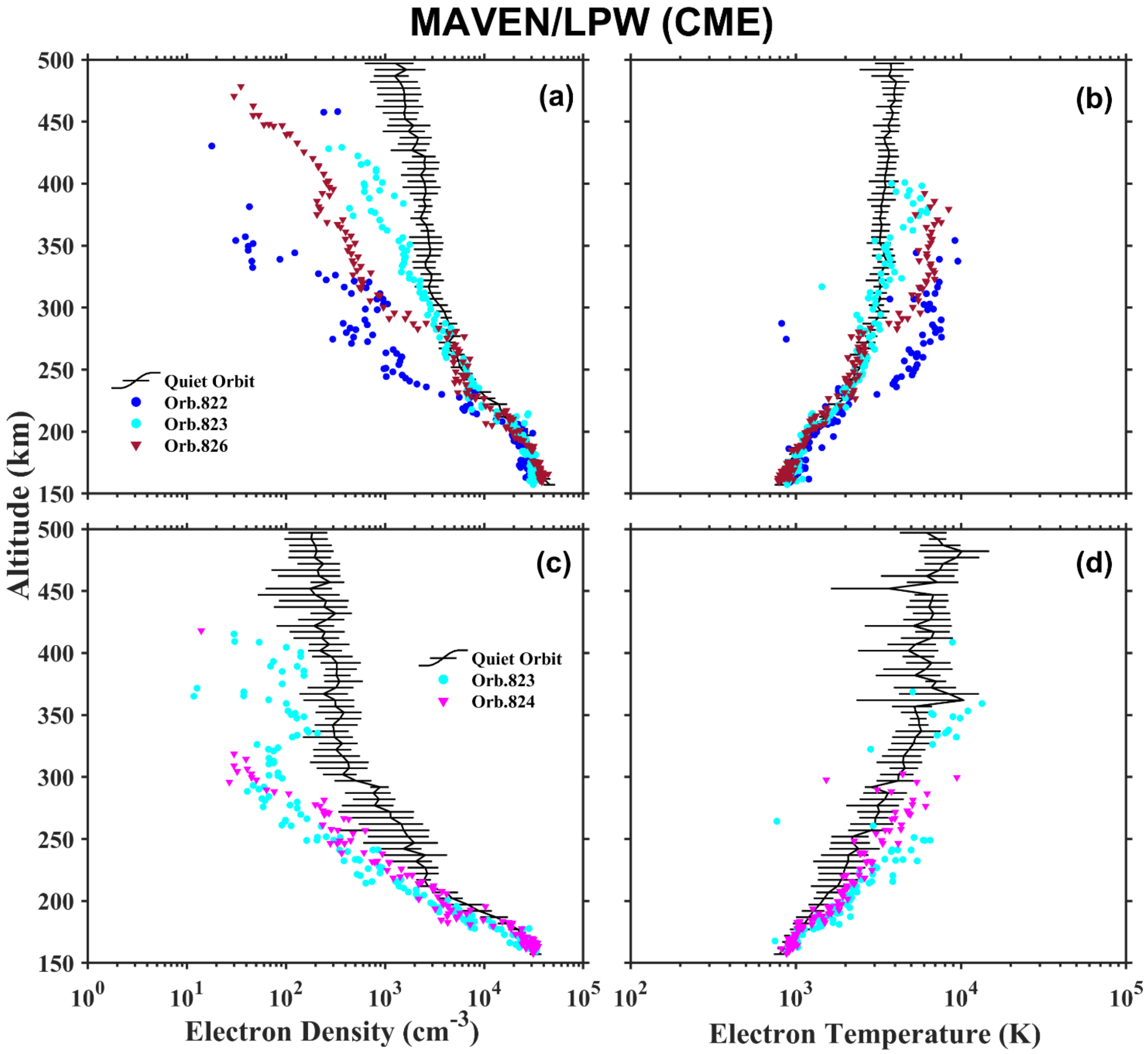

**Figure 4:** The Langmuir Probe and Waves (LPW), (a-b) dayside and (c-d) nightside, electron density and temperature variations as a function of altitude (150-500 km) during the 03-05 March 2015, along with the mean quiet-time variation. The disturbed orbit profiles are shown as different color dots and styles. The mean of the quiet-time profiles (five orbit profiles considered for both dayside and nightside, respectively) are shown (black curve) along with the standard deviation.

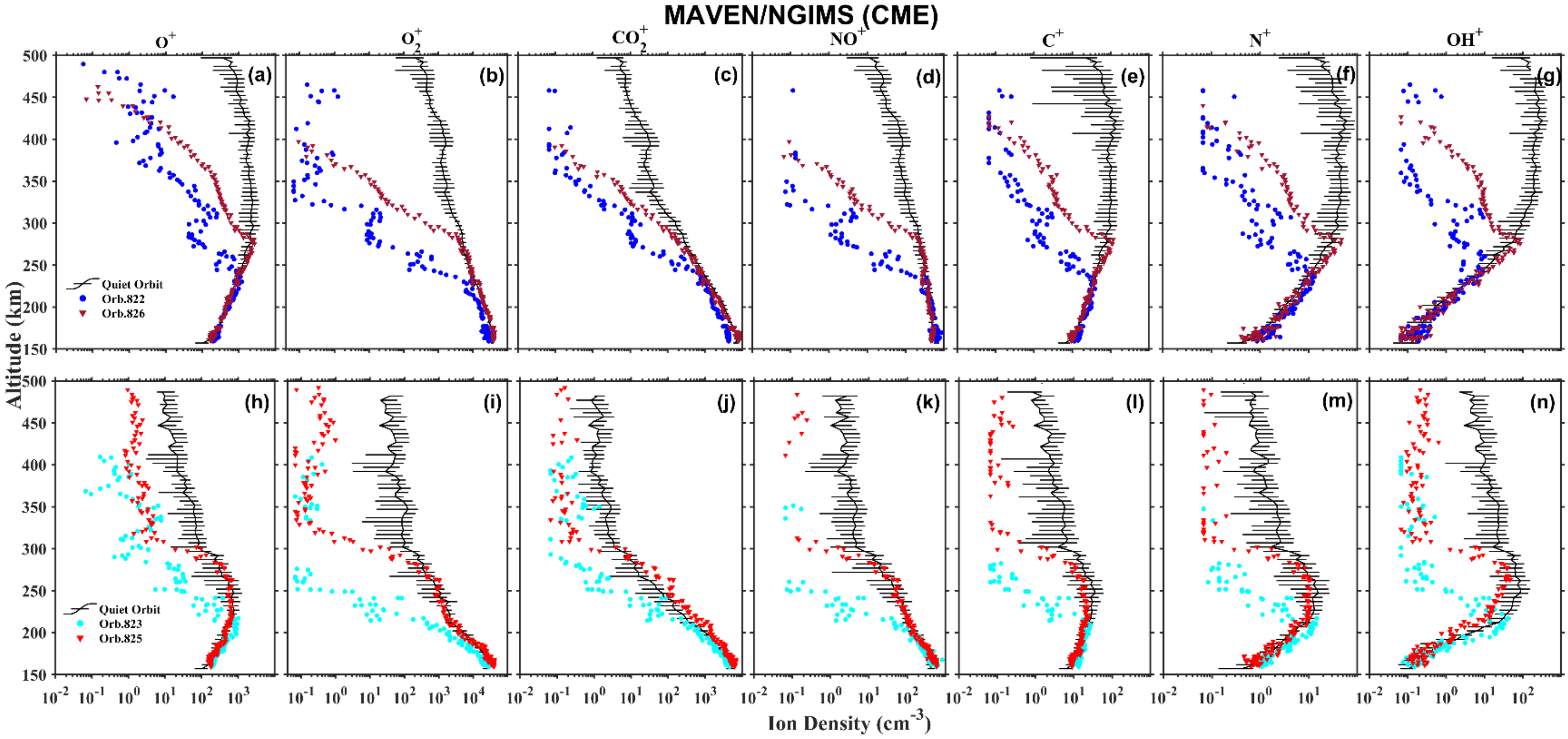

**Figure 5:** The Neutral Gas and Ion Mass Spectrometer (NGIMS), (a-g) dayside and (h-n) nightside, ions density variation as a function of altitude (150-500 km) during the 03-05 March 2015, along with the mean quiet-time variation. The disturbed orbit profiles are shown as different color dots and styles. The mean of the quiet-time profiles (five orbit profiles considered for both dayside and nightside, respectively) are shown (black curve) along with the standard deviation.

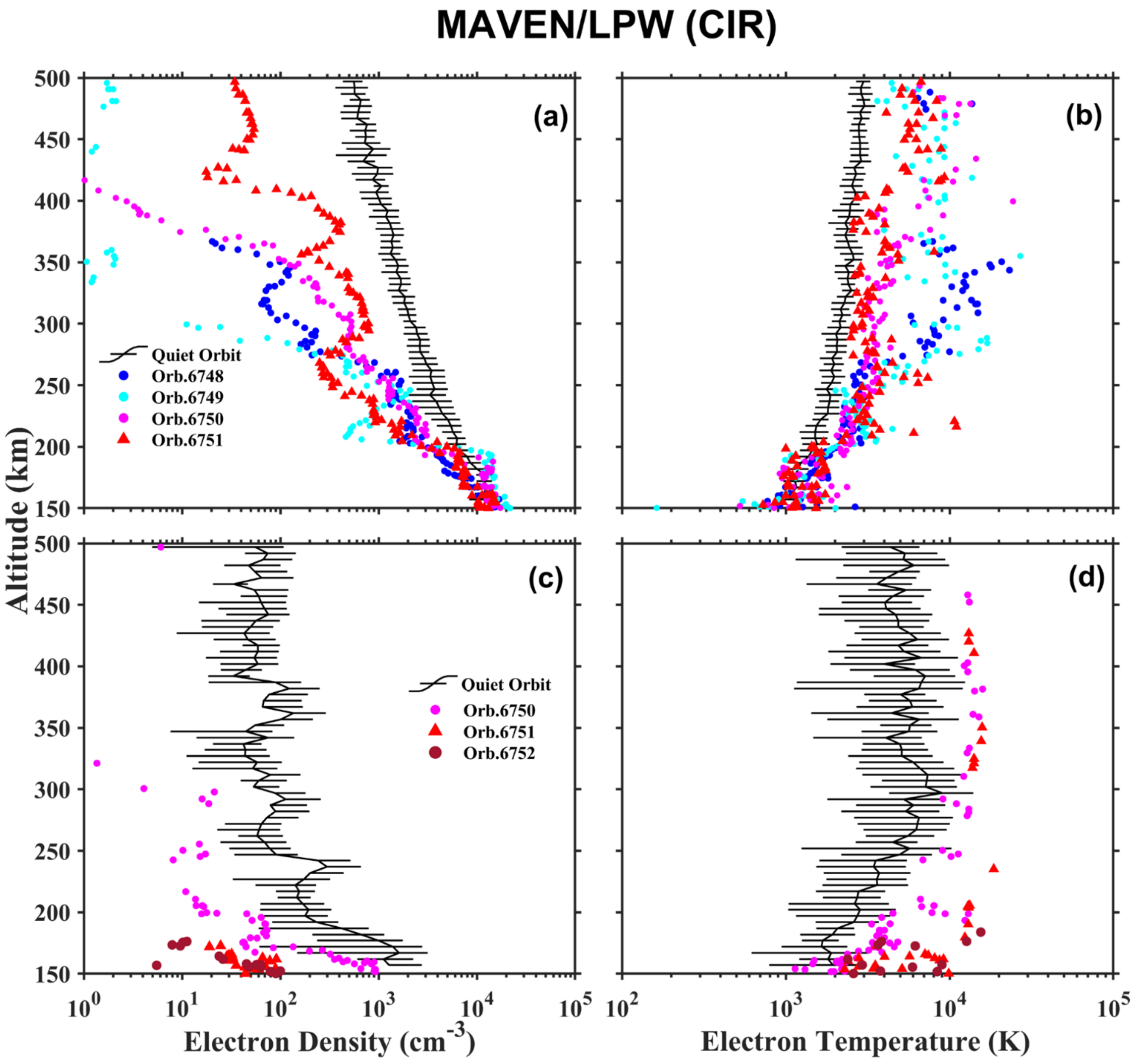

**Figure 6:** The Langmuir Probe and Waves (LPW), (a-b) dayside and (c-d) nightside, electron density and temperature variations as a function of altitude (150-500 km) during the 19-21 March 2018, along with the specific quiet-time variation. The disturbed orbit profiles are shown as different color dots and styles. The mean of the quiet-time profiles (five orbit profiles considered for both dayside and nightside, respectively) are shown (black curve) along with the standard deviation.

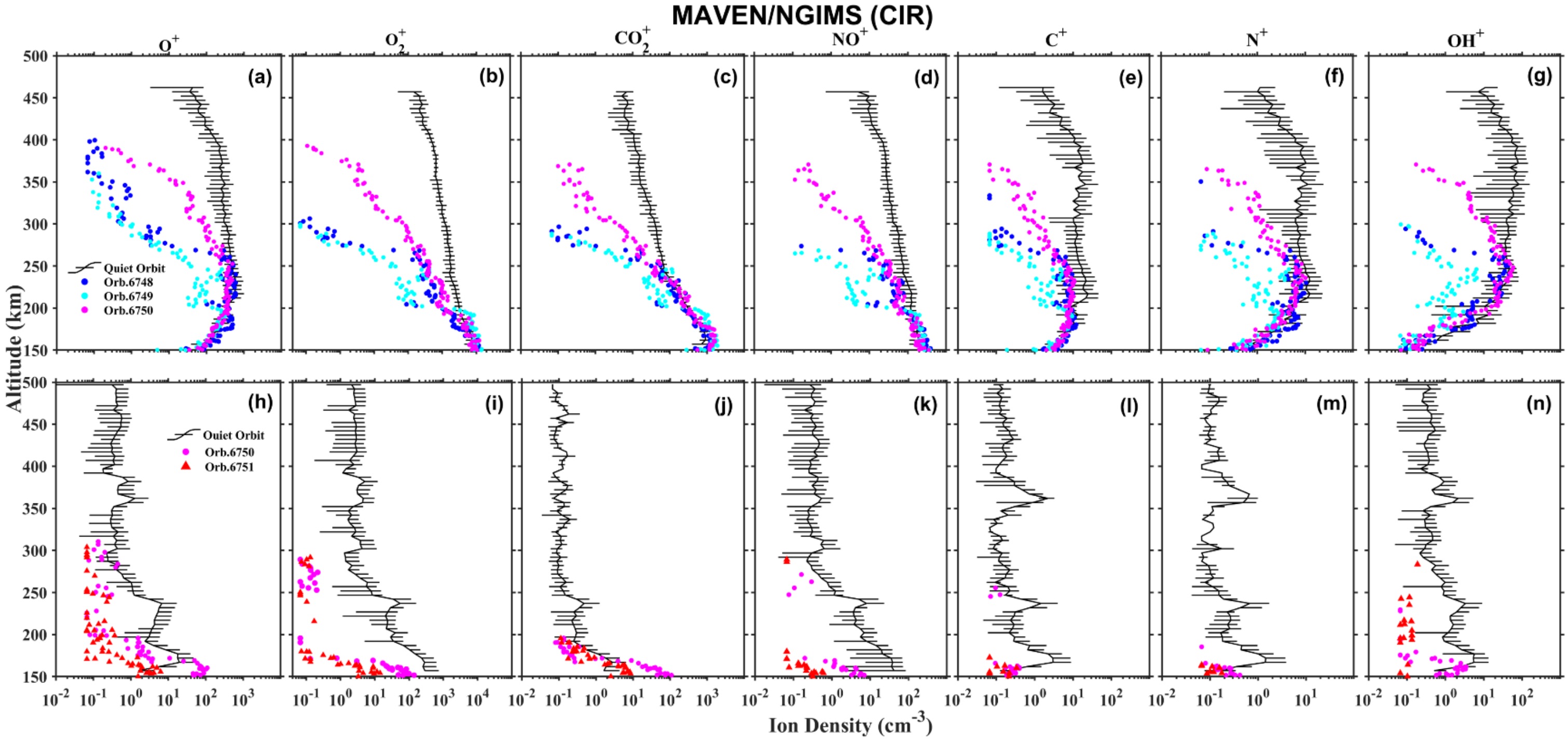

**Figure 7:** The Neutral Gas and Ion Mass Spectrometer (NGIMS), (a-g) dayside and (h-n) nightside, ions density variation as a function of altitude (150-500 km) during the 19-21 March 2018, along with the mean quiet-time variation. The disturbed orbit profiles are shown as different color dots and styles. The mean of the quiet-time profiles (five orbit profiles considered for both dayside and nightside, respectively) are shown (black curve) along with the standard deviation.

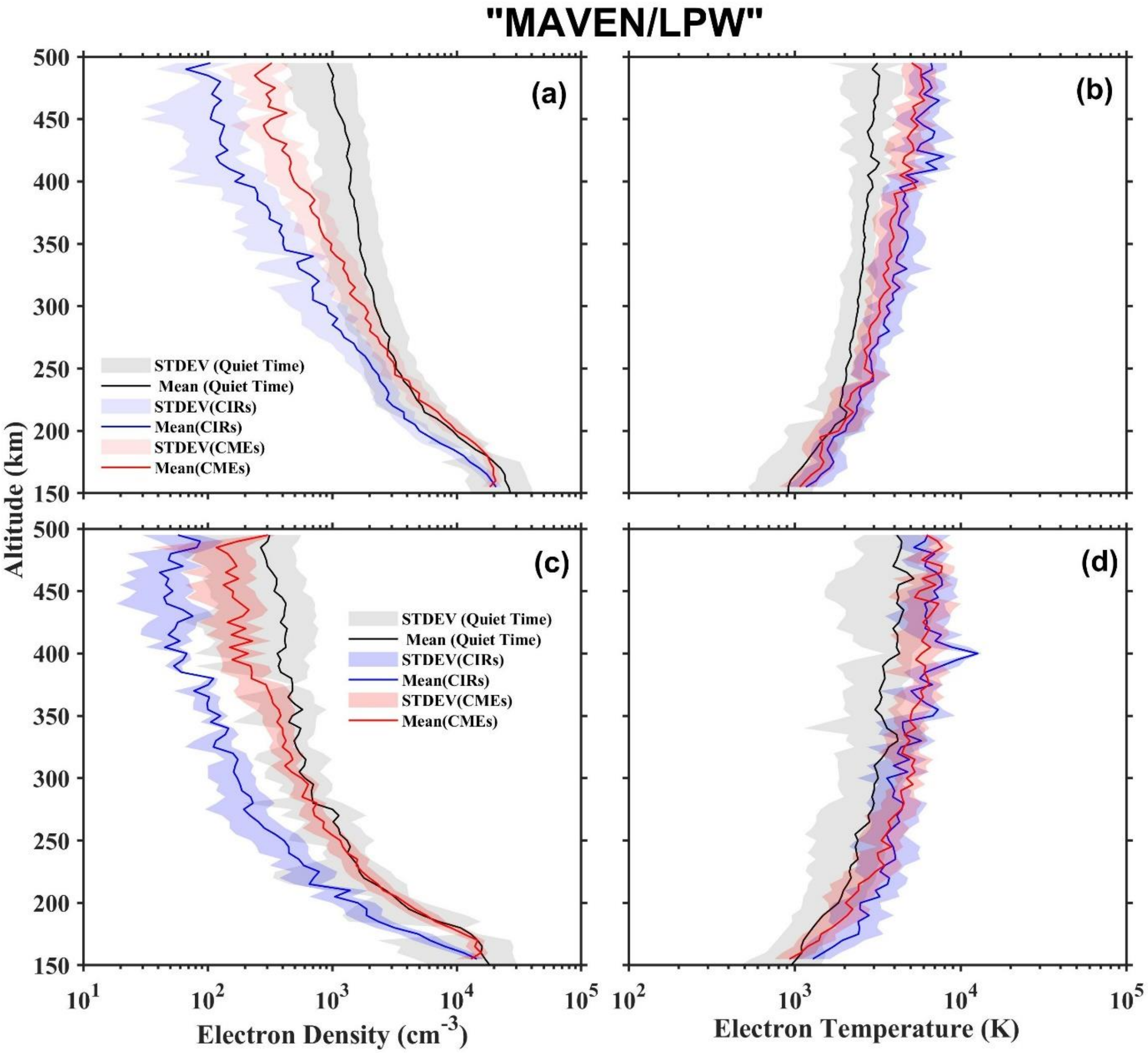

**Figure 8:** The Langmuir Probe and Waves (LPW), (a-b) dayside, (c-d) nightside, electron density and temperature variations as a function of altitude (150-500 km) for mean quiet-time orbit profiles (30 dayside and 30 nightside quiet-time orbit profiles), highly disturbed orbit profiles during 15 CIR events (55 dayside and 54 nightside disturbed orbit profiles) and 15 CME events (53 dayside and 55 nightside disturbed orbit profiles). The mean density and temperature represented as black, red, and blue profiles along with standard deviation (black, red and blue shades).

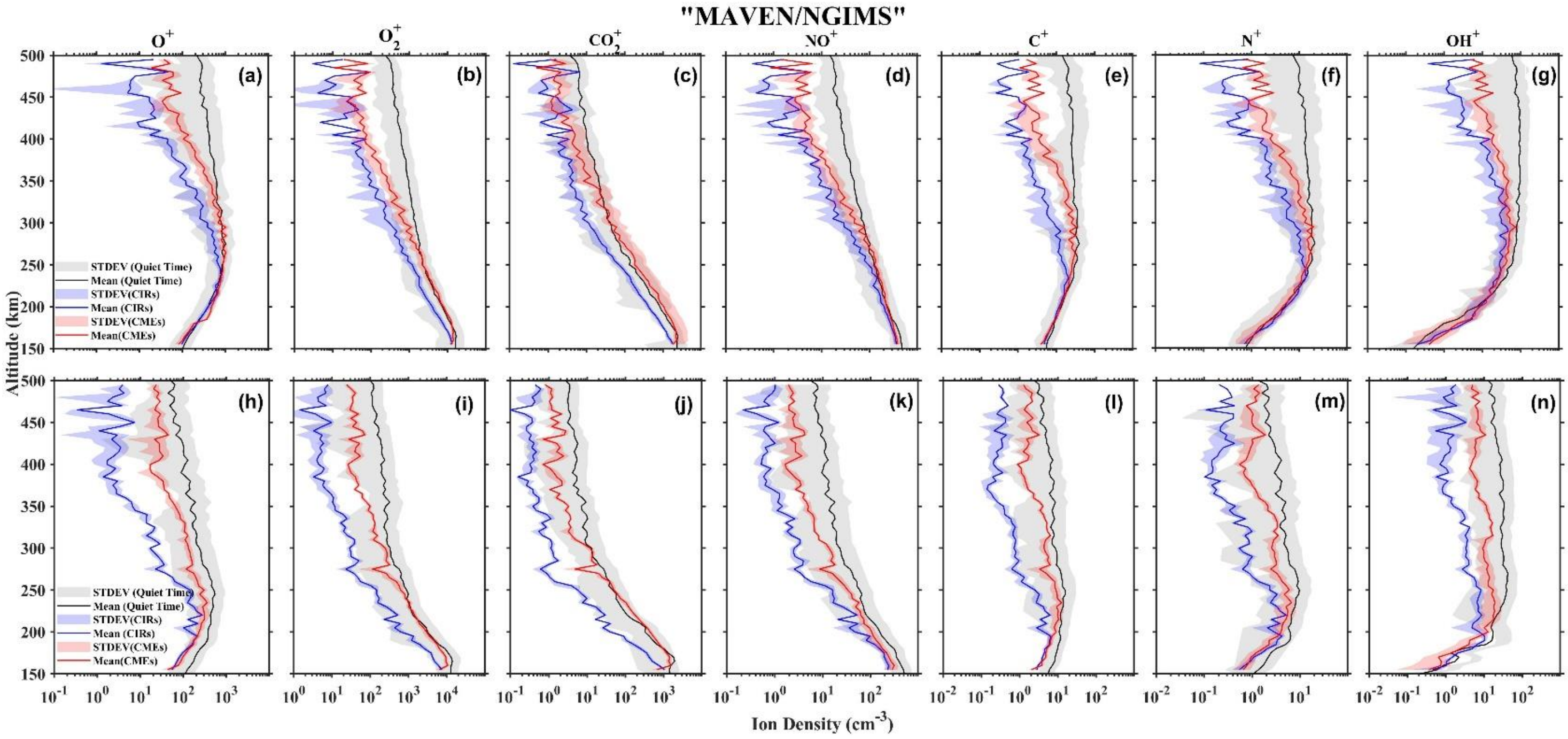

**Figure 9:** The Neutral Gas and Ion Mass Spectrometer (NGIMS), (a-g) dayside, (h-n) nightside, ions density variations as a function of altitude (150-500 km) for mean quiet-time orbit profiles (30 dayside and 30 nightside quiet-time orbit profiles), highly disturbed orbit profiles during 15 CIR events (55 dayside and 54 nightside disturbed orbit profiles) and 15 CME events (53 dayside and 55 nightside disturbed orbit profiles). The mean density and temperature represented as black, red, and blue profiles along with standard deviation (black, red and blue shades).

**Tables:**

| | S. No. | Start Date (YYYY/MM/DD) | End Date (YYYY/MM /DD) | No. of Disturbed Orbits | | LST (Hours) Dayside | SZA (°) Dayside | LST (Hours) Nightside | SZA (°) Nightside |
|---|---|---|---|---|---|---|---|---|---|
| | | | | Day side | Night side | | | | |
| CMEs | **1.** | 2015/02/17 | 2015/02/19 | 3 | 3 | 17.06 – 17.75 | 76 - 98 | 18.51 – 22.74 | 112-127 |
| | **2.** | 2015/02/25 | 2015/02/27 | 4 | 4 | 16.20 – 17.15 | 64 - 94 | 17.20 – 20.30 | 94 - 122 |
| | **3.** | *2015/03/03 | 2015/03/05 | 3 | 2 | 15.69 – 16.54 | 53 - 84 | 16.56 – 18.90 | 84 - 115 |
| | **4.** | 2015/03/08 | 2015/03/11 | 2 | 3 | 15.18 - 16.02 | 45.5 - 75 | 16.04 – 18.09 | 76 - 108 |
| | **5.** | 2015/10/04 | 2015/10/06 | 3 | 4 | 11.68 - 12.51 | 23 - 62 | 13.02 - 18.33 | 71 - 109 |
| | **6.** | 2016/09/23 | 2016/09/24 | 3 | 3 | 14.57 - 19.01 | 60 - 91 | 19.12 - 20.09 | 92 -120 |
| | **7.** | 2016/10/05 | 2016/10/07 | 4 | 2 | 11.78 - 17.77 | 52 – 72 | 17.79 – 18.91 | 72 - 99 |
| | **8.** | 2017/09/08 | 2017/09/11 | 3 | 4 | 16.61 – 17.70 | 68 - 73 | 17.90 – 0.26 | 73 - 87 |
| | **9.** | 2018/03/10 | 2018/03/11 | 3 | 3 | 15.00 – 18.20 | 77 - 105 | 22.50 – 1.50 | 123 -146 |
| | **10.** | 2018/04/09 | 2018/04/10 | 4 | 6 | 13.02 - 14.08 | 33 - 76 | 16.50 - 22.40 | 78 -122 |
| | **11.** | 2018/09/22 | 2018/09/25 | 4 | 3 | 12.62 – 20.00 | 90 -125 | 20.41 – 22.01 | 131-154 |
| | **12.** | 2018/10/03 | 2018/10/05 | 3 | 3 | 11.26 – 17.85 | 82 - 111 | 19.47 – 21.04 | 120 - 140 |
| | **13.** | 2018/10/12 | 2018/10/14 | 4 | 5 | 10.06 – 15.23 | 80 - 104 | 17.90 – 20.08 | 113 -129 |
| | **14.** | 2019/05/07 | 2019/05/12 | 6 | 4 | 11.82 – 14.57 | 61 - 79 | 04.34 – 07.13 | 91 -112 |
| | **15.** | 2020/03/04 | 2020/03/08 | 4 | 6 | 10.40 – 11.94 | 26.5 - 70 | 02.30 – 09.90 | 74 - 121 |
| | | | | | | | | | |
| CIRs | **1.** | 2015/09/03 | 2015/09/04 | 3 | 3 | 14.33 - 15.35 | 50.2 - 80.5 | 18.85- 23.96 | 107 - 135 |
| | **2.** | 2015/09/09 | 2015/09/11 | 3 | 4 | 13.80 - 14.73 | 44.2 - 80.2 | 17.50- 23.50 | 102 - 132 |
| | **3.** | 2015/12/17 | 2015/12/19 | 5 | 3 | 06.00 – 06.90 | 87 - 94 | 04.96 – 06.00 | 81 - 87 |
| | **4.** | 2016/03/11 | 2016/03/15 | 2 | 3 | 13.00 – 20.00 | 39.5 - 82 | 21.00 – 23.04 | 87 -130 |
| | **5.** | 2016/03/16 | 2016/03/19 | 2 | 2 | 12.00 – 19.00 | 35 - 78 | 20.00 – 22.00 | 82 - 125 |
| | **6.** | 2016/09/17 | 2016/09/22 | 3 | 4 | 16.16 – 19.75 | 65 - 97 | 19.75 – 20.67 | 98 - 128 |
| | **7.** | 2017/09/01 | 2017/09/02 | 3 | 3 | 17.30 – 18.50 | 75 - 78 | 18.50 – 01.72 | 80 - 88 |
| | **8.** | 2017/09/10 | 2017/09/11 | 3 | 3 | 16.50 – 17.50 | 66 - 70 | 17.66 – 23.77 | 73 - 86 |
| | **9.** | 2017/09/21 | 2017/09/22 | 5 | 5 | 15.50 – 16.50 | 56 – 59 | 16.52 – 20.70 | 59 -78 |
| | **10.** | 2018/03/09 | 2018/03/10 | 2 | 4 | 15.61 – 18.50 | 77 - 107 | 22.81 - 01.60 | 123 - 146 |
| | **11.** | *2018/03/19 | 2018/03/21 | 4 | 3 | 12.29 – 19.00 | 61.5 - 96 | 20.83 – 23.60 | 112 - 144 |
| | **12.** | 2018/09/22 | 2018/09/23 | 3 | 6 | 13.00 – 18.25 | 90 - 126 | 19.70 – 22.27 | 131 -154 |
| | **13.** | 2018/10/03 | 2018/10/06 | 7 | 3 | 11.26 – 17.85 | 82 - 111 | 19.50- 21.03 | 122 -140 |
| | **14.** | 2018/10/07 | 2018/10/11 | 6 | 5 | 10.80 – 17.00 | 80 - 108 | 18.50 - 20.60 | 118 - 135 |
| | **15.** | 2018/10/11 | 2018/10/18 | 4 | 3 | 10.20 – 15.64 | 80 - 105 | 18.00- 20.13 | 111 - 130 |

**Table 1:** MAVEN periapsis pass (during 15 CMEs and 15 CIRs), start date, end date, number of disturbed orbits, local solar time (LST), solar zenith angle (SZA) during the disturbed orbits in the Martian dayside and nightside ionosphere (measurement between 150 and 500 km altitude). *It represents the specific cases of CME and CIR events.

| Events | Dayside/ Nightside | [$O^+$] Mean peak Alt with STDEV (km) | [$O^+$] Mean peak Dens with STDEV ($cm^{-3}$) | [$C^+$] Mean peak Alt with STDEV (km) | [$C^+$] Mean peak Dens with STDEV ($cm^{-3}$) | [$N^+$] Mean peak Alt with STDEV (km) | [$N^+$] Mean peak Dens with STDEV ($cm^{-3}$) | [$OH^+$] Mean peak Alt with STDEV (km) | [$OH^+$] Mean peak Dens with STDEV ($cm^{-3}$) |
|---|---|---|---|---|---|---|---|---|---|
| **CMEs** | **Dayside** | 265±2 | 1017±165 | 255±2 | 31.15±7.2 | 265±2 | 17.44±4.1 | 295±2 | 71.73±22.12 |
| | **Nightside** | 235±2 | 361.85±78.8 | 235±2 | 12.62±2.98 | 235±2 | 6.69±1.89 | 235±2 | 19.27±9.14 |
| **CIRs** | **Dayside** | 235±2 | 810.8±122.4 | 235±2 | 21.18±4.8 | 235±2 | 13.95±2.61 | 290±2 | 47.50±20.56 |
| | **Nightside** | 225±2 | 276.6±39.25 | 220±2 | 6.28±0.86 | 220±2 | 5.20±0.68 | 190±2 | 10.52± 1.37 |

**Table 2:** The mean peak altitude and density with standard deviation for lighter ions ($O^+$, $C^+$, $N^+$, & $OH^+$) in the Martian dayside and nightside ionosphere during 15 CMEs and 15 CIRs. Alt, Dens and STDEV represent the Altitude, Density and Standard Deviation respectively.